%% file: main.tex
\documentclass{aa}
\usepackage{makecell}
\usepackage{comment}
\usepackage{booktabs}
\usepackage{graphicx}
\usepackage{color}
\usepackage{xcolor}
\usepackage{soul}
\usepackage{hyperref}
\usepackage{amsmath}
\usepackage{longtable}
\usepackage{txfonts}
\usepackage{subcaption}
\usepackage{multirow}
\usepackage[normalem]{ulem}

\begin{document} 
\newcommand{\rev}[1]{#1} 
\newcommand{\del}[1]{#1} 

   \title{EWOCS-XI: Anisotropic expansion and  shear in the cluster Westerlund\,1. 
   A 2D kinematic analysis with Gaia DR3}
    \titlerunning{Wd1}
   \author{L. Prisinzano
   \inst{\ref{ipris}}$^{\href{https://orcid.org/0000-0002-8893-2210}{\includegraphics[scale=0.5]{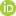}}}$\corrauth{loredana.prisinzano@inaf.it}
          \and
          F. Damiani
          \inst{\ref{ipris}}\email{francesco.damiani@inaf.it}
           \and
          C. Ordenes-Huanca{\inst{\ref{iordea},\ref{iordeb}}}\email{ccordenes@uc.cl}
          \and
           M. Tarantino
          \inst{\ref{itara}}\email{marco.tarantino01@unipa.it}
          \and
           S. Sciortino\inst{\ref{ipris}}\email{salvatore.sciortino@inaf.it}
          \and
          N.J. Wright
          \inst{\ref{iwrig}}\email{n.j.wright@keele.ac.uk}
          \and
          V. Almendros-Abad\inst{\ref{ipris}}\email{victor.almendrosabad@inaf.it}
          \and
          M. Andersen\inst{\ref{ieso}}\email{morten.andersen@eso.org}
          \and
          A. Bayo\inst{\ref{ieso}}\email{abayo@eso.org}
          \and
          R. Bonito\inst{\ref{ipris}}\email{rosaria.bonito@inaf.it}
          \and
          V. Cusimano\inst{\ref{icusi},\ref{ipris}}\email{vincenzo.cusimano@inaf.it}
          \and
          M. Haberle\inst{\ref{ieso}}\email{maximilian.haberle@eso.org}
          \and
          K. Mu\v{z}i\'c\inst{\ref{imuzi}}\email{kmuzic@ciencias.ulisboa.pt}
          \and
          I. Negueruela\inst{\ref{inegu}}\email{ignacio.negueruela@ua.es}
          \and 
          E. Sabbi\inst{\ref{isabbi}}\email{elena.sabbi@issibern.ch}
        \and
          A. Winter\inst{\ref{iwint}}\email{andrew.winter@qmul.ac.uk}
          \and
          M.G. Guarcello\inst{\ref{ipris}}\email{mario.guarcello@inaf.it}
          }
   \institute{INAF-Osservatorio Astronomico di Palermo, Piazza del Parlamento 1, 90134, Palermo,Italy\label{ipris}
   \and
   Instituto de Astrofísica, Pontificia Universidad Católica de Chile, Casilla 306, Santiago 7820436, Chile\label{iordea}
          \and
   Departamento de Astronomía, Universidad de Concepción, Casilla 160-C, Concepción, Chile\label{iordeb}
   \and
  Dipartimento di Scienze Economiche, Aziendali e Statistiche,
Universit\`a degli Studi di Palermo, Palermo, Italy\label{itara}
      \and
Astrophysics Group, Keele University, Keele ST5 5BG, UK\label{iwrig}
   \and
   European Southern Observatory, Karl-Schwarzschild-Strasse 2,
85748 Garching bei München, Germany\label{ieso}
  \and
  Dipartimento di Fisica e Chimica, Università degli Studi di Palermo, Palermo, Italy\label{icusi}
  \and
  Instituto de Astrof{\'\i}sica e Ci{\^e}ncias do Espa{\c{c}}o, 
Faculdade de Ci{\^e}ncias, Universidade de Lisboa, 
Ed. C8, Campo Grande, 
1749-016 Lisbon, Portugal\label{imuzi}
\and
Departamento de Física Aplicada, Facultad de Ciencias, Universidad de Alicante, Carretera de San Vicente s/n, 03690 San Vicente
del Raspeig, Spain\label{inegu}
\and
   Gemini Observatory/NSFNOIRLab, 950N.CherryAve., Tucson, AZ85719, USA\label{isabbi}
   \and
   Astronomy Unit, School of Physics and Astronomy, Queen Mary University of London, London E1 4NS, UK\label{iwint}
   }
   \date{Received 24 July 2026 ; accepted 6 October 2026}

\abstract
  {Westerlund 1 (Wd1) is one of the most massive young clusters in the Milky Way, serving as a fundamental laboratory for studying extreme star formation and massive stellar evolution. However, its heavy extinction 
  and the high Galactic field star contamination
  have historically hampered the precise characterization of its stellar population and internal dynamics in the optical bands.}
  {We aim to identify the members of Wd1 with Gaia DR3 to provide a publicly accessible catalog. Furthermore, we seek to investigate the internal kinematics of the cluster to assess its current dynamical state, determine the influence of the Galactic potential, and provide insights  on its formation scenario.}
  {We leveraged the enhanced astrometric precision of \textit{Gaia} Data Release 3 (DR3). By applying a robust 5D clustering approach, we identified the bona fide cluster members. We then modeled their proper motions using a 2D kinematic first-order Taylor expansion to derive the velocity gradient tensor, enabling a kinematic tomography across the cluster to map expansion, rotation, and shear.}
  {We successfully isolated a robust sample of 559 Wd1 members. To ensure 
maximum astrometric reliability, our subsequent kinematic analysis was restricted 
to a subset of 524 stars satisfying a total proper motion uncertainty  
of $< 0.5~\mathrm{mas\,yr^{-1}}$.
At least 103 of them are classified as gravitationally unbound members, including 14 are walkaway stars. 
  The 2D kinematic analysis reveals a statistically significant divergence (>4$\sigma$) with preferred expansion along the Galactic latitude direction with a position angle of $ 164^\circ\pm\rev{10.1}^\circ$. We also detect a  cross-gradient 
  ($\partial v_b / \partial l = -0.41 \pm 0.08$ km s$^{-1}$ pc$^{-1}$), which provides a clear kinematic signature of a shear on the plane of the sky.
}
  {Our analysis reveals that the cluster  does not behave as a single cohesive population, but rather consists of two kinematically decoupled subsystems: a stable, gravitationally bound core in dynamical equilibrium and an escaping, unbound envelope. \rev{We detect a prominent anisotropic expansion perpendicular to the Galactic plane. The inferred internal kinematic expansion timescale ($\Delta t_{\mathrm{kin}} = 2.08 \pm 0.44$\,Myr) matches the independent orbital mid-plane crossing time ($\Delta t_{\mathrm{orbit}} = 2.10 \pm 0.05$\,Myr). This temporal alignment indicates that the onset of the expansion
   happened during the cluster's Galactic disk passage,
  though its physical trigger could stem either from internal dynamical processes (e.g., early supernova feedback) or from external Galactic tidal perturbations.}
Finally, the identification of a robust population of unbound members provides  evidence that a cluster dissolution process is actively underway, likely driven by discrete dynamical ejection events rather than  long-term stellar evaporation.}
  
   \keywords{  
   Stars: massive --
                Galaxy: open clusters and associations: individual: Westerlund 1 --
                Methods: data analysis --
                Catalogs
               }

\maketitle
\nolinenumbers
%
\section{Introduction \label{sec:intro}}
Westerlund\,1 (Wd1) is a cornerstone for our understanding of star formation in extreme environments.
It is widely considered the most massive young cluster in the Milky Way, with total mass estimates ranging from $\approx 4 \times 10^4$ to over $10^5\,\mathrm{M}_{\odot}$ \citep{clar05, bran08, ande17}, where the primary uncertainties arise from the assumed cluster distance, isochrone age, and high-mass IMF slope.
 Its high density of massive stars, including Wolf-Rayet stars, yellow hypergiants, and red supergiants, makes it a crucial laboratory for testing theories of high-mass stellar evolution and binary interaction \citep{clar11,anas24}, as well as their effects in the cloud dispersal and the characteristics of the intermediate and low-mass population. Wd1 is the target of the multiwavelength campaign EWOCS (Extended Westerlund 1 and 2 Open Clusters Survey) which is primarily based on observations obtained with JWST, Chandra, ALMA and HST \citep{guar24}. 

In a recent  study, \citet{negu22} utilized Gaia Early Data Release 3 (EDR3) and AAT/AAOmega spectroscopy to characterize the cluster and its environment. They derived a robust distance of $4.23^{+0.23}_{-0.21}$ kpc and identified an extinction wall at approximately 2.8 kpc, likely associated with molecular clouds in the Scutum-Crux arm. Furthermore, they confirmed the presence of a foreground population at $d \approx 2$ kpc, which they linked to the open cluster BH 197. Their membership selection, based on the  non-parametric analysis of proper motions, identified 401 bona fide members but noted significant field contamination and incompleteness in the cluster core due to extreme crowding.

\rev{\citet{nava22} independently evaluated Wd1's distance using Gaia EDR3, highlighting the historical difficulties in estimating distances due to severe and variable extinction. By strictly restricting their analysis to 5-parameter astrometric solutions to avoid unreliable parallax zero-point corrections for highly reddened stars, they derived a distance of $4.06_{-0.34}^{+0.36}$\,kpc. Crucially, they validated this Gaia-based result with a completely independent, non-astrometric technique, i.e. through the geometric modelling of the eclipsing binary W36, which yielded $4.03 \pm 0.25$\,kpc \citep{roch22}. This modern convergence securely places the cluster at $\sim 4.1 - 4.2$\,kpc. Establishing a stable distance framework is critical also for our study, as it directly impacts the reliable conversion of proper motions into the physical internal velocities required to evaluate the cluster's kinematics.}


\rev{Despite these advances, the extraordinary complexity of the Wd1 field, characterized by extreme and spatially variable reddening, requires careful consideration. While earlier studies often relied on standard extinction assumptions yielding $A_V \approx 10-12$ (e.g., Negueruela et al. 2010), Damineli et al. (2016) demonstrated that Wd1 follows a significantly steeper, non-standard reddening law with $R_V = 2.50 \pm 0.04$. Furthermore, they derived an average extinction of $A_V = 11.40 \pm 2.40$ and revealed a severe spatial extinction gradient across the cluster face spanning $\Delta A_V \sim 8.7$. This highly variable extinction, combined with overlapping Galactic structures, complicates traditional photometric analyses, making a high-precision, wide-field Gaia DR3 kinematic analysis of its cluster both timely and crucial. }

 The first aim of this work is to build upon and  extend the census provided by \citet{negu22} to provide the astronomical community with a robust, publicly accessible catalog of \textit{Gaia} members to facilitate future multi-wavelength investigations. 
To achieve this, we exploit the enhanced astrometric precision of \textit{Gaia} Data Release 3 (DR3) 
\citep{prus16,vall23,brow21}, 
alongside a robust 5D clustering approach to distinguish the cluster's population from the dense Galactic background. In addition, by leveraging \textit{Gaia} DR3 proper motion precision, we aim to investigate the internal kinematics of Wd1 in order to assess its current dynamical state and its interaction with the Galactic potential.

While near-infrared (NIR) observations are crucial to pierce through the heavy interstellar extinction of Wd1 and detect its highly obscured stellar population
\rev{\citep[e.g.][]{bran08,genn11,genn17,ande17,orde26a,orde26b}},
Gaia DR3 optical observations offer an unparalleled astrometric precision for kinematic studies. When sensitivity is not a limiting factor, Gaia's proper motions provide a cleaner and more precise mapping of the cluster's internal dynamics. Ultimately, both wavelength regimes and datasets present complementary advantages and limitations for the study of such extreme systems.

The paper is structured as follows. Section~\ref{sec:data} describes the \textit{Gaia} DR3 data retrieval and the initial sample selection. 
Section~\ref{sec:methods} focuses on the membership selection and its validation. 
Section~\ref{sec:kinematics_results} is dedicated to the 2D kinematic modeling of Wd1, where we derive its macroscopic internal kinematics. In Section~\ref{sec:discussion}, we discuss our findings, comparing them with recent near-infrared studies to evaluate the dynamical state and formation scenario of the cluster. Finally, Section~\ref{sec:conclusions} summarizes our main conclusions.

\section{Data and sample selection}
\label{sec:data}
\begin{figure*}[t]
   \centering
\includegraphics[width=\textwidth]{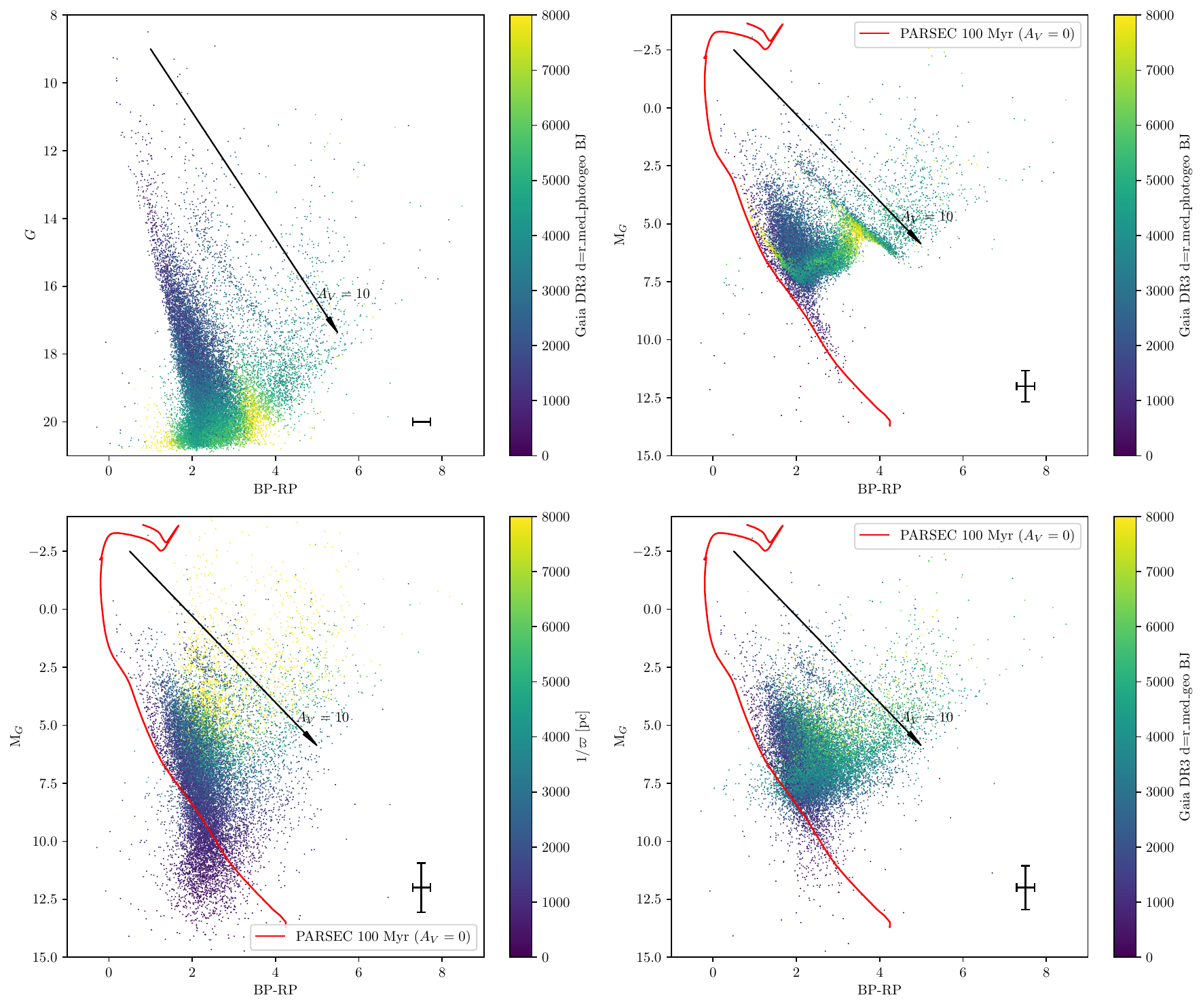}
   \caption{Color-Magnitude Diagrams for the Wd 1 field
   (top-left). 
            Absolute magnitude $M_G$ vs. observed color $(BP-RP)$ using  
            \texttt{r\_med\_photogeo} (top-right),
            $1/\varpi$ (bottom-left), 
            \texttt{r\_med\_geo} (bottom-right) distances. Color-coding represents the distance adopted in each panel. 
            The red line indicates a 100 Myr solar metallicity PARSEC isochrone  at zero extinction \citep{bres12}. Representative error bars are shown in each panel.}
   \label{fig:camd_comparison}
\end{figure*}
The primary dataset for this study was retrieved from the Gaia DR3 archive. We performed a cone search centered on the nominal coordinates of the Westerlund 1 cluster ($\alpha = 251.7678^\circ, \delta = -45.857^\circ$) with a radius of $13'$,
to cover available surveys. 

The data were acquired via an ADQL query\footnote{https://www.ivoa.net/documents/ADQL/20231215/index.html} using the \texttt{astroquery.gaia} Python interface. We performed a direct cross-match between the main \texttt{gaiadr3.gaia\_source} table and the \texttt{external.gaiaedr3\_distance} catalog \citep{bail21} using the unique \texttt{source\_id} identifier. This cross-match provided us with both the standard \textit{Gaia} photometry and astrometric parameters (positions, proper motions, and parallaxes) as well as the high-precision geometric and photo-geometric distances required to trace the 3D structure of the region.

No preliminary quality cuts (e.g., on magnitude limits or parallax signal-to-noise ratio) were applied. This strategy ensures the maximum possible completeness, preventing the unintended exclusion of genuine, highly reddened massive members of Wd1, which might otherwise be filtered out by strict astrometric constraints prior to our subsequent multidimensional clustering analysis.
    The selected catalog contains 23762 unique sources within the target area. 

The errors on the Gaia DR3 $G$, $G_{BP}$, and $G_{RP}$  magnitudes were computed as:
\begin{equation}
e_{mag} = \sqrt{\left( \frac{2.5}{\ln(10)} \cdot \frac{e_{flux}}{flux} \right)^2 + \sigma_{0}^2}
\end{equation}
where $\sigma_{0}$ represents the zero-point uncertainties: $\sigma_G = 0.002755$, $\sigma_{BP} = 0.002790$, and $\sigma_{RP} = 0.003779$. 
The zero-point offsets were taken from the ESA Gaia Mission documentation\footnote{\url{https://www.cosmos.esa.int/web/gaia/edr3-passbands}}. 

\section{Methods}
\label{sec:methods}
\subsection{Distance Calibration and Absolute Magnitudes}
\label{sec:photometry}

We evaluated three different distance indicators for each source: the direct inversion of the parallax ($d = 1/\varpi$), and the median geometric (\texttt{r\_med\_geo}) and photo-geometric (\texttt{r\_med\_photogeo}) distances provided by \citet{bail21}. 

Figure~\ref{fig:camd_comparison} presents a multi-panel diagnostic comparison between the observed Color--Magnitude Diagram (CMD; top-left panel) and absolute magnitude representations (CAMD; remaining panels). In the top-left panel, stars are color-coded by their photo-geometric distance (\texttt{r\_med\_photogeo}). The other three panels show the resulting CAMDs obtained using different distance estimators: the photo-geometric distance (\texttt{r\_med\_photogeo}; top-right), direct parallax inversion ($1/\varpi$; bottom-left), and the geometric distance (\texttt{r\_med\_geo}; bottom-right).

Overplotted in red is an unreddened PARSEC isochrone (100~Myr, $A_V = 0$) by \citet{bres12} \del{. to illustrate the impact of each distance prescription on the dispersion of the stellar sequences.}
\rev{We emphasize that the observed stellar magnitudes used to construct these CAMDs have not been corrected for interstellar extinction; they are apparent magnitudes converted to absolute magnitudes ($M_G$) relying solely on the respective distance modulus. The unreddened $A_V=0$ isochrone is plotted here not to fit the cluster, but to act as a zero-extinction reference sequence for the abundant foreground Main Sequence (MS) field stars. This allows us to visually evaluate which distance indicator best aligns this largely unreddened foreground population with the theoretical stellar locus. As expected, the heavily obscured Wd1 sequence is physically displaced from this reference isochrone by the extreme interstellar reddening along the line of sight (illustrated by the $A_V=10$ vector).}

These CAMDs are derived using the entire catalog without applying any preliminary photometric or astrometric quality cuts. This inclusive approach is strictly necessary to retain the actual population of Wd1; given the cluster's large distance and extreme extinction, its members are faint in the optical bands and inherently possess large relative astrometric uncertainties. For instance, enforcing a standard quality filter, such as requiring a relative photo-geometric distance error smaller than 20\%, would systematically discard most of the bona fide cluster members, severely biasing the sample toward the unreddened foreground population. 

We find that the photo-geometric distances (\texttt{r\_med\_photogeo}) provide the most consistent alignment with a theoretical unreddened 100 Myr  isochrone (used here to represent the unreddened foreground \del{Main Sequence, } MS), significantly reducing the vertical scatter caused by astrometric uncertainties in the high-reddening regime of Wd1. 

Based on this comparison, in this paper we adopted the photo-geometric distances as the primary distance indicator for the subsequent membership analysis and population classification. We emphasize that because these distances are mainly used to build the CAMD and identify the cluster photometric region, the impact of retaining sources with large relative distance errors on our final results is small.
 Crucially, individual distance uncertainties do not significantly propagate into or bias the subsequent spatial and kinematic cluster membership (Sect. 3.3), which relies mainly on high-precision 2D proper motions.
\subsection{Photometric Population Selection}
\label{sec:selection}
\begin{figure*}[h]
   \centering
\includegraphics[width=0.95\textwidth]{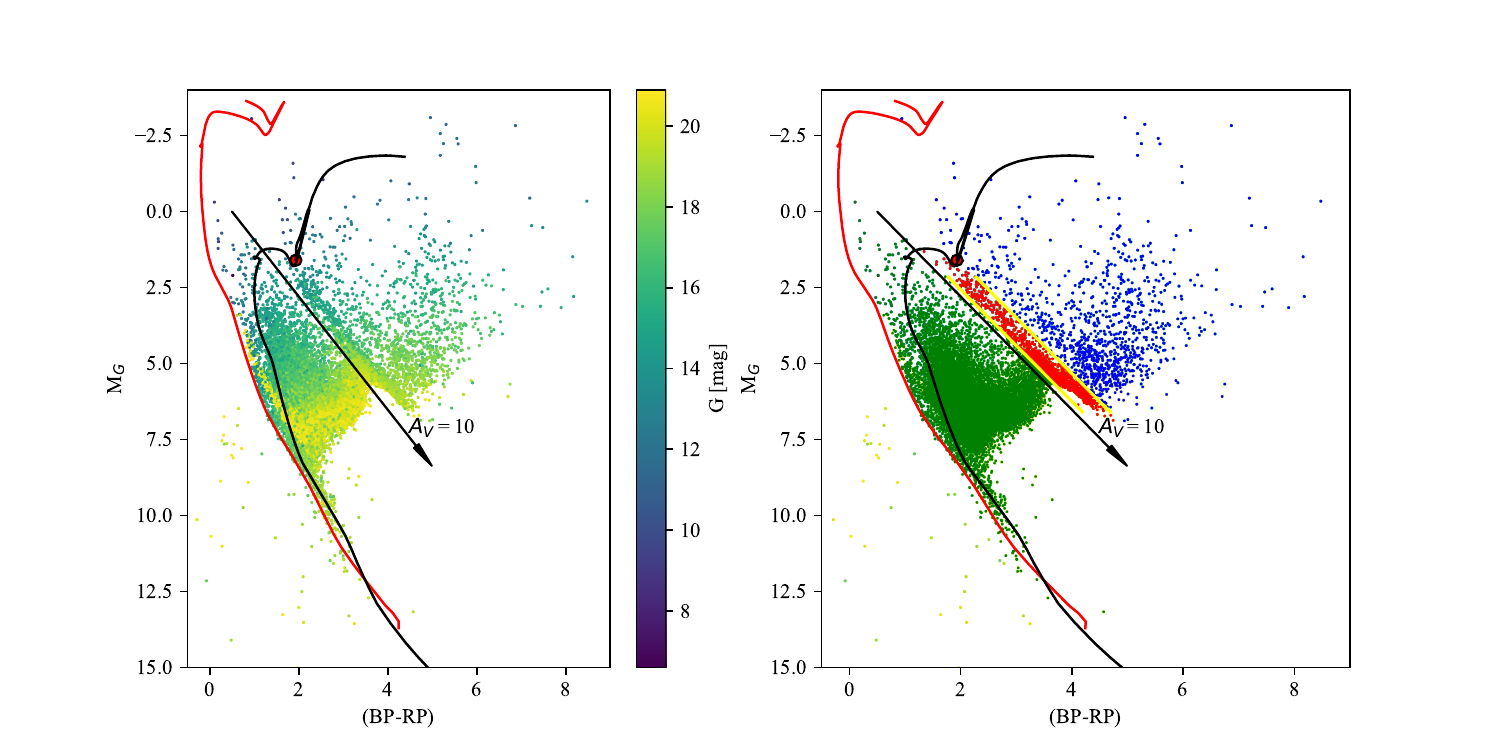}
   \caption{\rev{CAMD of the stellar populations in the Wd1 field without applying extinction corrections. 
   \textit{Left panel}: Diagram $M_G$ vs. $BP-RP$ for all Gaia DR3 sources, color-coded by apparent $G$ magnitude. The solid red line represents an unreddened 100~Myr PARSEC isochrone ($A_V=0$). The solid black line shows a 500~Myr PARSEC isochrone at $A_V=2.2$, with the red dot highlighting the intrinsic theoretical locus of the RC. The black arrow indicates the reddening vector for $A_V=10$. 
   \textit{Right panel}: Same diagram showing the results of the photometric classification. Blue points represent the Wd1 photometric candidate members, red points identify the  candidate RC giants, and green points show the  field population. Yellow lines delimit the empirical selection boundaries used to separate the RC giants obtained by translating the intrinsic RC locus along the reddening vector.}            }
\label{fig:photo_selection}
\end{figure*}

To disentangle the diverse stellar populations overlapping with Wd1 along the line of sight, we performed a multi-stage selection in the  CAMD shown in  Fig.\,\ref{fig:photo_selection}.  The diagram shows clearly the \del{Main Sequence (MS)} \rev{MS} population affected by relatively low reddening, as can be deduced from the comparison with the overplotted  100 Myr old PARSEC isochrone at zero extinction by \citet{bres12}.

In addition, the CAMD exhibits a clear sequence of Red Clump (RC) giants affected by strong reddening. 
To select this population, we first identified the RC evolutionary phase from the 500\,Myr model-predicted isochrone, which corresponds to a nominal stellar mass of $2.574\ M_\odot$ at the onset of the core helium-burning stage.

Using the Gaia DR3 adopted extinction law\footnote{following the prescription given in \url{https://www.cosmos.esa.int/web/gaia/edr3-extinction-law}},
we calculated the model-predicted coordinates of this RC phase at two reference extinction limits: a minimum of $A_0 = 2.2$ mag and a maximum of $A_0 = 9.0$ mag. 
These specific limits were chosen to empirically match the minimum and maximum reddening boundaries of the observed Galactic RC sequence along this line of sight.
We then defined a mean reddening track by linearly connecting these two extreme points in the magnitude-color space. Finally, to encompass the observed data spread, we constructed a selection strip by shifting this track empirically along the reddening vector direction, applying a tolerance of $-0.1$ mag to define the blue boundary and $+0.4$ mag for the red boundary.

The  blue boundary \del{($-0.1~\mathrm{mag}$)} provides a strict threshold to minimize contamination from unreddened or moderately reddened foreground field stars. Conversely, the broader red margin \del{($+0.4~\mathrm{mag}$)} accounts for the observed stellar density, likely due to combined broadening effects of localized differential extinction and unresolved stellar multiplicity. For clarity, the $A_V = 10~\mathrm{mag}$ reddening vector is displayed in both panels of Fig.~\ref{fig:photo_selection}.

Then, we defined as candidate Wd1 members the 1091 
objects occupying the upper-right region of the CAMD, brighter than the RC limit. This region is physically expected to host the most luminous and massive population of the cluster. Indeed, while the extreme interstellar extinction ($A_V \approx 10 - 12~\mathrm{mag}$) characterizing this line of sight systematically shifts the entire cluster sequence—from low-mass stars up to massive supergiants—to redder colors and fainter magnitudes \citep[e.g.,][]{bran08, genn11, ande17}, our magnitude threshold brighter than the RC limit specifically isolates the upper \del{main sequence} \rev{MS} and evolved massive stars.

Finally we defined as foreground Field stars all sources 
redder than the MS isochrone and occupying the lower-left region of the CAMD, fainter than the RC limit (green dots in 
Fig.\,\ref{fig:photo_selection}, right panel).

The selected photometric samples were used  for the subsequent 5D astrometric clustering analysis described below.

\subsection{Astrometric Membership and Clustering Analysis}
\label{sec:clustering}
\begin{figure*}[ht]
   \centering
   \includegraphics[width=0.95\textwidth]{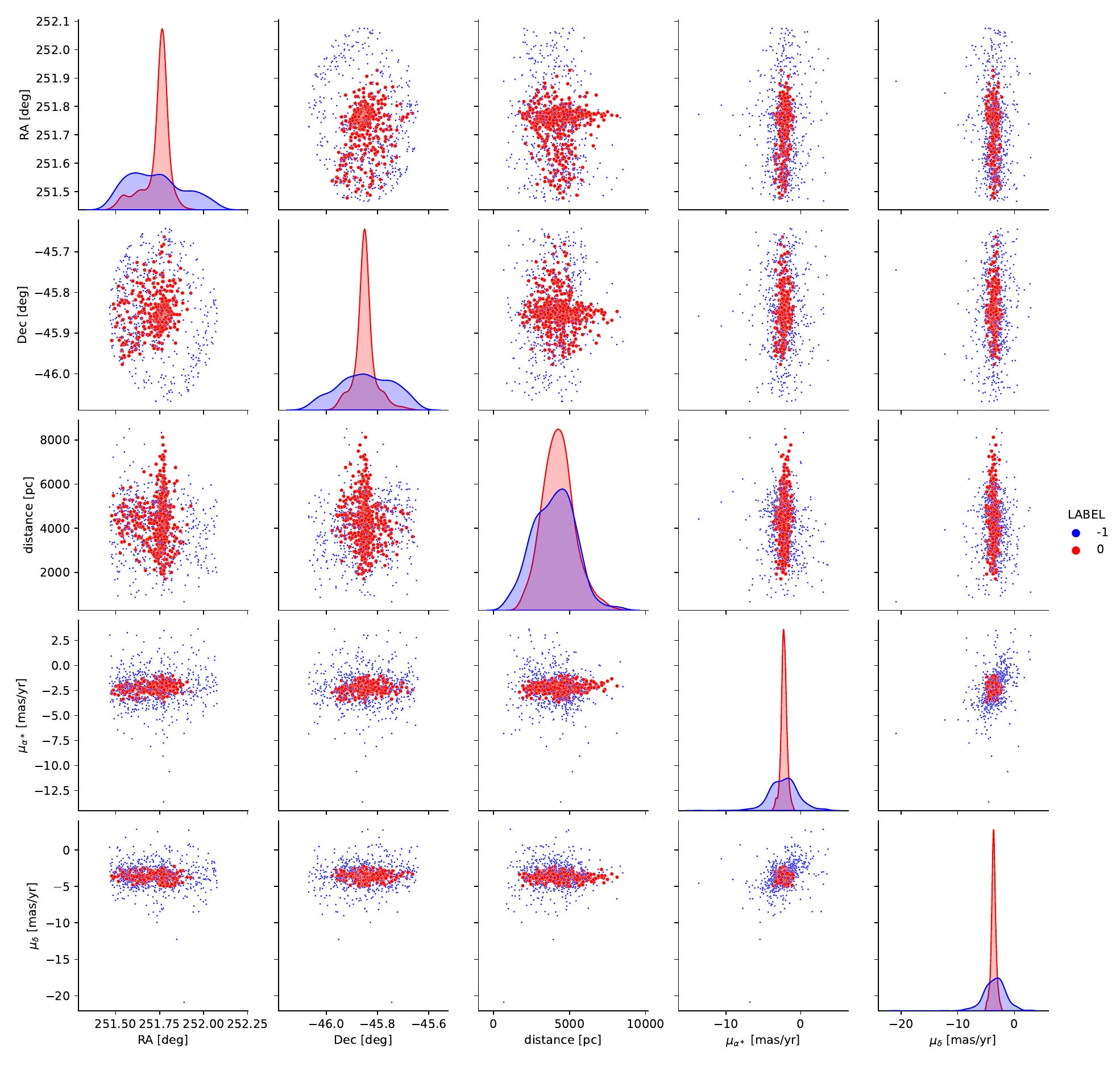}
   \caption{Astrometric membership validation of Wd1 using the DBSCAN clustering algorithm. The pair plot displays the multidimensional distribution of the photometrically selected stars in the 5D astrometric space, including equatorial coordinates ($\alpha, \delta$), photo-geometric distances, and proper motions ($\mu_{\alpha\ast}, \mu_{\delta}$). Red symbols identify the high-probability cluster members. Blue symbols represent the discarded field stars. The Kernel Density Estimation (KDE) plots are shown along the diagonal.}
   \label{fig:dbscan_results}
\end{figure*}
To isolate the physical members of Wd1 from the dense Galactic foreground and background populations, we employed the \textit{Density-Based Spatial Clustering of Applications with Noise} (DBSCAN) algorithm. Unlike traditional selection methods, DBSCAN identifies clusters of arbitrary shape based on the local density of points in a multidimensional parameter space \citep{este96}.

For this analysis, we considered the sample of 1091 photometrically selected Wd1 candidate members and we defined a 5D astrometric space consisting of
celestial coordinates ($\alpha, \delta$), to account for spatial concentration,
proper motions ($\mu_{\alpha*}, \mu_{\delta}$), to identify stars with coherent kinematic signatures,
and photo-geometric distances from \citet{bail21}.

Prior to clustering, the data were pre-processed using a \texttt{RobustScaler} \citep{pedr11}. This scaler normalizes each of the five astrometric dimensions independently by subtracting its median to normalize the features, mitigating the influence of outliers and ensuring that each astrometric dimension contributed equitably to the Euclidean distance calculation.

We performed an optimization of the DBSCAN hyperparameters, specifically the maximum neighbor distance ($\epsilon$) and the minimum number of points required to form a cluster (\textit{min\_samples}). Through a grid search, we identified the optimal configuration at $\epsilon=0.9$ and \textit{min\_samples}=10 for the primary Wd1 selection.

The results, illustrated in the diagnostic pair plot in Fig. \ref{fig:dbscan_results}, show the clear isolation of a primary cluster (\texttt{LABEL=0}) composed of 559 high-probability members. These stars exhibit a tight distribution in the proper motion plane ($\mu_{\alpha}$ and $\mu_{\delta}$)
and are concentrated at a median distance of $\sim 4213$\,pc, with a median absolute deviation (MAD) of $\sim 659$\,pc \del{. This observed distance spread is entirely dominated by individual measurement uncertainties, as the true physical depth of the cluster remains unresolved along the line of sight using the Gaia DR3 parallax errors. 
However, thanks to the large number of selected members, the derived median value provides a well-constrained and reliable estimate of the cluster's distance.}
\rev{(corresponding to a standard deviation of $\sigma \approx 977$~pc), which is entirely dominated by
 individual Gaia DR3 parallax measurement noise rather than true physical depth. 
 To derive the distance uncertainty of the cluster, we separate the individual distance dispersion 
 from the uncertainty on the cluster centroid.
The formal statistical standard error of the median cluster distance is given by 
$\sigma_{\rm stat} \approx 1.253 \times {\rm MAD} / \sqrt{N} \approx 35$~pc for $N = 559$ members.
The dominant source of uncertainty is systematic, arising from the residual spatial variations 
of the Gaia DR3 parallax zero-point offset. \citet{maiz22} derived a minimum systematic parallax 
uncertainty of $\sigma_{\varpi,\rm sys} = 10.3~\mu{\rm as}$  for individual stars 
and compact clusters in Gaia EDR3/DR3.
At a cluster parallax of $\varpi \approx 0.238$~mas ($d \approx 4.21$~kpc), 
this systematic uncertainty of $\sigma_{\varpi,\rm sys} \approx 0.0103$~mas 
translates to a systematic distance error of $\sigma_{\rm sys} \approx 0.18$~kpc ($\approx 4.3\%$).
Combining the statistical and systematic uncertainties in quadrature 
($\sigma_d = \sqrt{\sigma_{\rm stat}^2 + \sigma_{\rm sys}^2} = \sqrt{0.035^2 + 0.18^2} \approx 0.19$~kpc), 
and accounting for spatial prior calibration uncertainties in the photo-geometric distance model 
\citep{bail21}, we \rev{derive} a final cluster distance of $d = 4.21 \pm 0.20$~kpc.}
%
%
%
%

 The remaining 532 sources not meeting the DBSCAN density criteria were classified as field noise (\texttt{LABEL=-1}) and excluded from the final member list.
This multidimensional approach ensures a robust census of the cluster population, minimizing contamination from the field.

To check and compare our density-based selection, we additionally performed an independent clustering analysis using the Hierarchical Density-Based Spatial Clustering of Applications with Noise (HDBSCAN) algorithm. Unlike the flat clustering approach of DBSCAN, HDBSCAN extracts clusters from a hierarchical tree of densities. To ensure the algorithm did not artificially fragment the extended, lower-density halo of the cluster, we configured the model to extract finer-grained structures by adopting the \texttt{leaf} cluster selection method. The hyperparameters were set to a minimum cluster size (\textit{min\_cluster\_size}) of 15 and \textit{min\_samples}=5. Furthermore, we applied a \textit{cluster\_selection\_epsilon} of 0.1 to prevent contiguous kinematic structures from being split into micro-clusters. This selection strategy  yielded a larger, more inclusive sample of 738 candidate members.

To further investigate the discrepancies between the two clustering approaches, we performed a direct source-by-source cross-match between the respective candidate lists. The comparison reveals a perfect 100\% overlap with respect to the density-based selection: all 559 members identified by DBSCAN are entirely recovered by HDBSCAN. 
To validate our kinematic selections and assess the purity of the two samples, we cross-matched both the DBSCAN and HDBSCAN candidate lists with the comprehensive X-ray catalog of Wd1 published by \citet{guar24}. X-ray emission is a robust and independent tracer of stellar youth and activity, making it an excellent discriminant against older, quiescent field stars in the Galactic plane.

The cross-match revealed that 327 out of the 559 DBSCAN candidates (58.5\%) have an X-ray counterpart. In contrast, the HDBSCAN selection yielded 349 matches out of 738 candidates (47.3\%). Although the HDBSCAN algorithm identified 179 additional sources, only 22 of these exhibited X-ray emission. This drastic drop in the X-ray detection fraction among the additional sources strongly suggests that the more inclusive HDBSCAN extraction (the \texttt{leaf} method) incorporated a significant amount of kinematic field contamination.

In addition we note that while the fraction of X-ray sources in our DBSCAN sample of 559 members is 58.5\% (327/559), the fraction of  X-ray emitters among the stars discarded by DBSCAN  (selected as field stars) is 13\% (69/532). \del{This significant difference in the X-ray detection fraction (58.5\% vs 13\%) indicates that our DBSCAN selection is highly effective at isolating cluster members from the  Galactic field population.}
\rev{This high X-ray detection fraction 
(58.5\% among the DBSCAN-selected sources compared to 13\% in the remaining sample) strongly supports that our DBSCAN selection reliably isolates genuine, X-ray-active cluster members concentrated in the central region of Wd1.} Furthermore, this 58.5\% detection rate is  high considering that late-B and early-A type stars are known to be intrinsically very weak (if any) X-ray emitters at the age of Wd1.

\subsection{Membership Probabilities}

To assess the geometric and kinematic stability of our membership and assign empirical membership probabilities to each member, 
we developed and executed a rigorous Monte Carlo (MC) resampling simulation over $B = 1000$ independent realizations.

For each MC replica, the observed 5D parameters of the 1091 initial 
photometric candidates were stochastically perturbed within their formal Gaia 
DR3 error distributions. Coordinates ($\alpha, \delta$) and proper motions 
($\mu_{\alpha*}, \mu_{\delta}$) were resampled assuming independent Gaussian 
errors. Conversely, because standard symmetrical errors propagate poorly for 
distant sources, the photo-geometric distances were perturbed using an 
asymmetric, log-normal distribution defined by the lower and upper confidence 
limits ($r_{\text{lo}}$, $r_{\text{hi}}$) provided by \citet{bail21}. 
During each iteration, the perturbed 5D space was self-consistently 
re-normalized  and the 
DBSCAN algorithm  was 
re-executed as for the initial sample.

\begin{figure}
\centering
\includegraphics[width=\columnwidth]{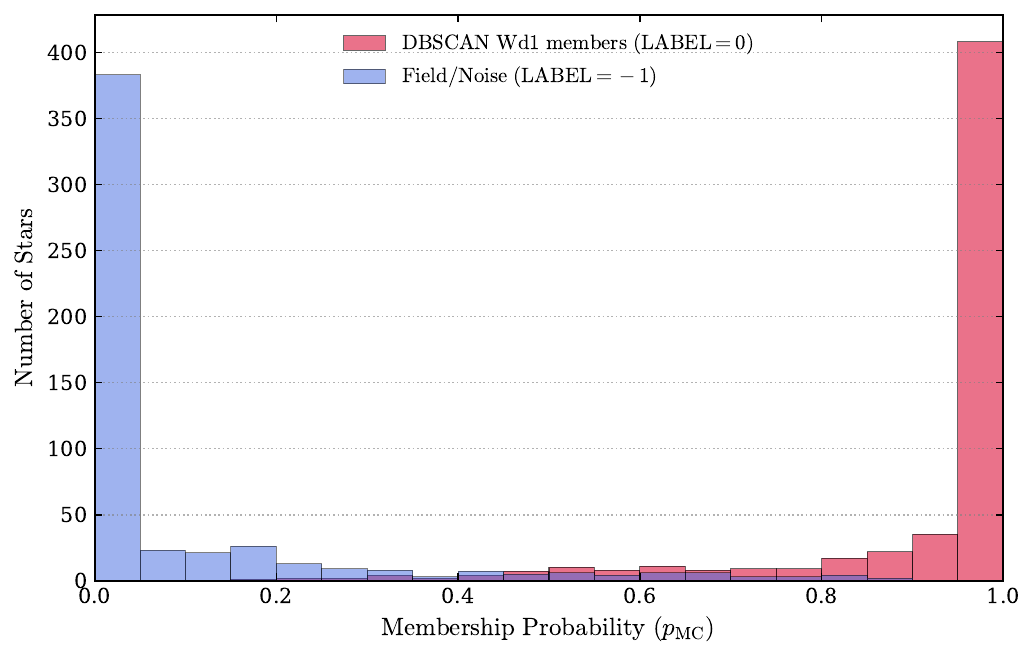}
\caption{Distribution of the empirical Monte Carlo membership probabilities 
for the Westerlund~1 candidate sample based on $B = 1000$ structural 
perturbation loops. The blue histogram displays the stability distribution 
for the 559 high-probability members isolated by the nominal, flat DBSCAN 
clustering algorithm ($\text{LABEL} = 0$). The red histogram shows the 
distribution for the 532 sources rejected as nominal field noise 
($\text{LABEL} = -1$). The sharp bimodal stratification confirms the high 
topological stability and low field contamination of the density-based 
selection boundaries.}
\label{fig:membership_histogram}
\end{figure}

The empirical membership probability $p_{\text{MC}}$ for each star was 
computed as the ratio between the number of successful recoveries in the 
matched Wd1 cluster and the total number of simulations ($B$). Subsequently, 
treating the cluster assignment as a binomial process, we assigned a Monte 
Carlo standard error to each probability value, defined as:
\begin{equation}
\sigma_{p} = \sqrt{\frac{p_{\text{MC}}(1 - p_{\text{MC}})}{B}}
\end{equation}
The resulting distribution of these empirical probabilities is illustrated in 
Fig.~\ref{fig:membership_histogram}. The separation between the two nominal DBSCAN populations is evident: out of the 559 nominal cluster 
members ($\text{LABEL} = 0$), the vast majority ($N = 443$, or $\sim 79\%$) 
exhibit $p_{\text{MC}} \ge 0.90$, and the 
entire sample displays a median probability of $0.995$. Conversely, the 532 
stars discarded as nominal field noise ($\text{LABEL} = -1$) are heavily 
concentrated at $p_{\text{MC}} = 0.00$, with a mean background probability of 
just $0.089$. 

The two distributions overlap only marginally within their extreme 
tails, with only 22  cluster members ($\sim 3.9\%$) displaying an 
uncertain membership probability of $p_{\text{MC}} < 0.50$. This minimal 
overlap highlights the stability and robustness of the 
DBSCAN selection.

Consequently, we conclude that the DBSCAN algorithm provides a better balance between completeness and purity. The highly reliable DBSCAN selection of 559 members is therefore adopted as our final cluster census for the subsequent analysis.
 The complete catalog of these selected members, including their \textit{Gaia} DR3 astrometric and photometric parameters, as well as the adopted photo-geometric distances from \citet{bail21}, is made publicly available in Table~\ref{tab:w1_catalog}.
\input{FIGURES/wd1_mem}

Following the characterization of the main cluster, we applied the same 
5D DBSCAN clustering methodology to the photometric sample associated 
with the foreground population, from which we identified the open cluster BH197 \citep{negu22}. 
The detailed analysis and stellar properties of this foreground population 
are given in Appendix~\ref{app:bh197}.

\subsection{Cluster membership validation}
\label{sect:validation}

Due to the large distance of Wd1 ($d \approx 4.21$~kpc) and the intrinsic limitations of the Gaia DR3 parallax uncertainties in the high-extinction regime, individual trigonometric parallaxes poorly constraints the DBSCAN selection. As a consequence, the algorithm primarily relied on the spatial concentration of the sources on the sky plane ($\alpha, \delta$) and their coherent proper motion signatures ($\mu_{\alpha*}, \mu_{\delta}$) to isolate the co-moving population of the cluster.

We note that to ensure that this density-based approach was not contaminated by random alignments of the dense Galactic background, the clustering analysis was executed on a strictly pre-selected photometric sample. By leveraging the Gaia photometry and 
constructing a stable CAMD  using the photo-geometric distances provided by \citet{bail21}, we pruned the master catalogue of the dominant field contaminants, excluding, for example, the unreddened foreground MS stars, the foreground cluster BH197 and the heavily populated, highly obscured field RC giant sequences.

To validate our final Wd1 member census, we cross-matched our sample (consisting of 524 stars with a total proper motion uncertainty threshold of $< 0.5$ mas~yr$^{-1}$) with \rev{three independent reference catalogues for this field: the spectroscopic survey of luminous stars by \citet{clar20} and the Gaia EDR3-based membership selection by \citet{negu22} and \citet{nava22}.}

To find the counterparts of the stars included in the \citet{clar20} catalogue within our list of cluster members, we used the empirical distribution of angular separations. To eliminate confusion from spurious matches, we used a conservative astrometric matching radius of $0.6^{\prime\prime}$. Our DBSCAN sample includes 121 successful matches out of the 166 (73\%) spectrally confirmed cluster members catalogued by \citet{clar20}.

To test our completeness in the brightest regime, we considered the subset of stars brighter than $RP = 13.5$ mag, an interval where the spectroscopic census of massive stars is expected to be largely complete \citep{negu22}. Within our catalogue, 72 stars satisfy this brightness threshold, yielding  46 successful matches with the \citet{clar20} catalogue. Remarkably, this represents a significant improvement compared to the 27 bright members structurally confirmed as cluster members by \citet{negu22}, proving that our approach successfully retrieves genuine cluster populations that were  rejected by \citet{negu22}.

To perform a homogeneous comparison with the Gaia EDR3 census by \citet{negu22}, we replicated their exact selection boundaries on our dataset, isolating stars with RUWE<1.4,  $(BP-RP) > 4.0$ mag and located within a projected radius of $3.5^\prime$ from the cluster's nominal center. This spatial and photometric filter isolates a sample of 359 stars.

By cross-matching this sub-sample with the 401 members listed in the catalogue of \citet{negu22}  we find 303 common stars.
The fact that a remarkably high fraction of our selected members (303/359, or $\sim$84\%) has a counterpart in an independent astrometric and photometric selection suggests that the DBSCAN algorithm isolates a highly pure population of stars that are physically consistent with the true stellar population of Westerlund~1.

\rev{To further evaluate the completeness of our membership selection against established massive stars, we cross-matched our catalog with the 97 confirmed Wd1 members possessing Gaia EDR3 astrometric solutions, found by \citet{nava22}. Both our DBSCAN and HDBSCAN selections recover 86 out of these 97 sources (88.7\%), confirming high completeness and strong consistency for the known massive stellar population of the cluster.}

\subsection{\del{Comparative Isochrone Fitting} \rev{Isochrone Comparison} and Population Validation}
\label{sec:evolution}
\rev{The final validation of the identified Wd1 population was performed by placing the selected members in the CAMD using the adopted fixed cluster distance ($d = 4.21$\,kpc) and comparing them with PARSEC isochrones (Bressan et al. 2012). }

\rev{We emphasize that a formal isochrone fitting to derive individual ages or stellar masses is unfeasible using optical photometry alone. Due to the extreme and highly variable interstellar extinction and the non standard   
reddening law 
toward Wd1 \citep[][Cusimano et al., submitted]{dami16}, the selected members do not form a single tight sequence in the CAMD, but stretch along the reddening vector.}

\begin{figure}
   \centering
      \includegraphics[width=\columnwidth]{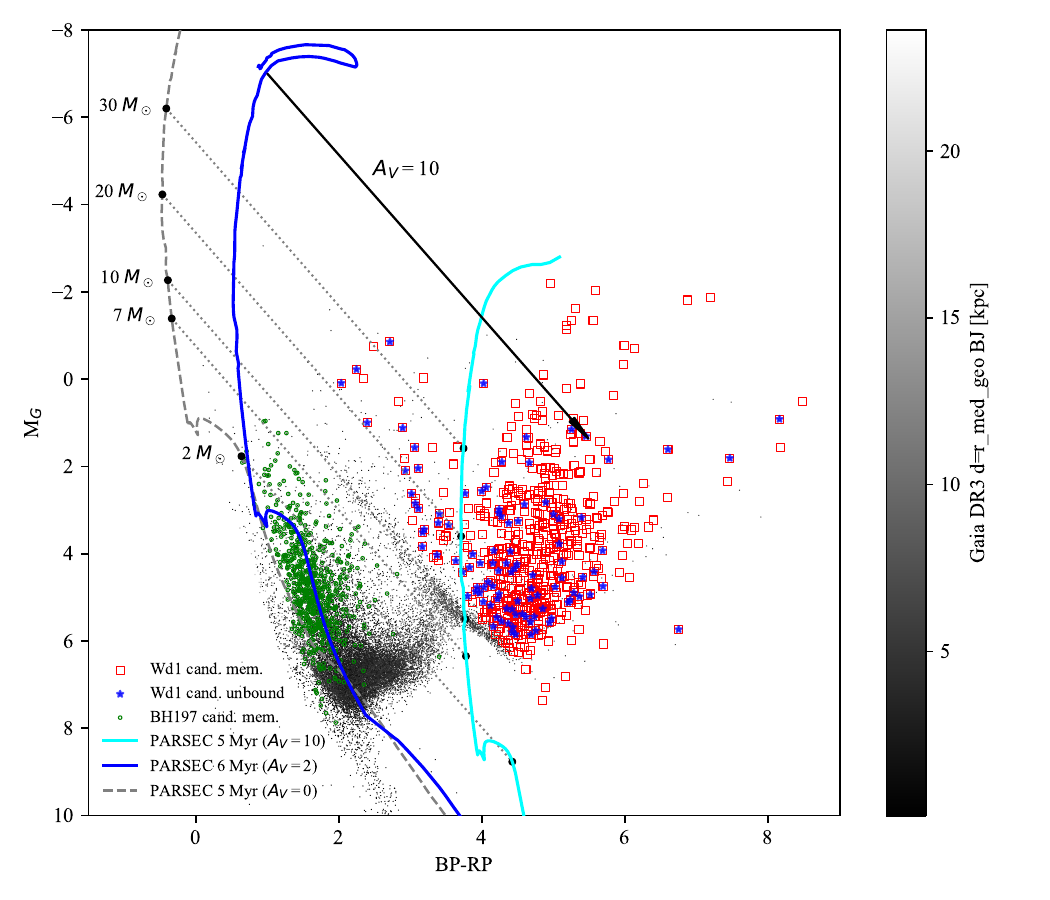}
   \caption{
   CAMD of the Wd1 field. The blue and cyan solid lines represent the PARSEC isochrones for an age of 6~Myr ($A_V = 2$~mag) and 5~Myr ($A_V = 10$~mag), respectively. The gray dashed line illustrates the intrinsic, unreddened 5~Myr isochrone ($A_V = 0$~mag). Selected initial stellar masses ($2$, $7$, $10$, $20$, and $30~M_\odot$) are highlighted as black dots along the unreddened sequence and are connected via dotted gray lines to their corresponding reddened positions on the $A_V = 10$~mag isochrone. The full field sample is plotted as a grayscale background color-coded by the Gaia~DR3 geometric distances ($r_{\rm med\_geo}$). Red squares and green circles denote the candidate members of Wd1 and BH197, respectively, as retrieved by the DBSCAN algorithm. 
   Blue x symbols are those selected as unbound candidate members (see Sect.\,\ref{sec:unbound}). 
   \rev{Wd1 members are plotted here adopting
   the fixed cluster distance of 4.21\,kpc.}
   The solid black arrow indicates the reddening vector corresponding to $A_V = 10$~mag, assuming a standard $R_V = 3.1$ extinction law converted into the Gaia EDR3/DR3 passbands.             }
   \label{fig:final_analysis}
\end{figure}

\rev{As shown in Fig.~\ref{fig:final_analysis}, a 5\,Myr PARSEC isochrone with $A_V = 10$ \citep{bres12} represents a lower envelope for the cluster sequence. 
Adopting $A_V \approx 10$ as a baseline minimum extinction is in  agreement with \citet{dami16}, who determined a mean total extinction of $A_V = 11.40 \pm 2.40$ toward Wd1, with an interstellar foreground component of $A_V \approx 9.7$ and a strong spatial reddening variation of $\Delta A_V \sim 8.7$ across the cluster.}

\rev{Despite these photometric limitations, overplotting theoretical sequences demonstrates that our optical selection specifically isolates the upper MS and evolved massive stars ($M \gtrsim 7\,M_\odot$).}

\del{The final characterization of the identified clusters was performed by comparing the results in the CAMD uncorrected for reddening with the solar metallicity PARSEC \mbox{\citep{bres12}} isochrones. As shown in Fig.~\ref{fig:final_analysis}, while the foreground cluster BH197 is qualitatively reproduced by a $6\,$Myr isochrone with a uniform, moderate extinction ($A_V \approx 2$), the isochronal analysis for Wd1 is much harder due to the high extinction. In fact, the selected members of Wd1 do not trace a tight, well-defined sequence in the CAMD. Instead, they exhibit a massive spread in color (over 4 magnitudes in $BP-RP$) and absolute magnitude.}

\del{We adopt a PARSEC isochrone of $5\,$Myr \mbox{\citep{cast26}}, covering a mass range between 0.1 and 32\,$M_\odot$, and a nominal extinction of $A_V=10$ 
\mbox{\citep[][Cusimano et al., submitted]{negu10}}.} 
\del{Although it intersects part of the population, the isochrone fails to encompass the entire distribution, meaning that $A_V=10$ is likely a lower limit to the cluster reddening. This large spread is a direct consequence of extreme and highly variable  reddening across the cluster core, coupled with the  limitations of the large \textit{Gaia} DR3 parallax uncertainties at the cluster distance.Consequently, characterizing  ages and individual masses for the Wd1 members purely from optical photometry is currently unfeasible.
Considering that the bulk of cluster members are in the $M_G$ range between 0 and 6.5\,mag, assuming the adopted isochrone, we roughly estimate that the selected members have masses between 7 and $\gtrsim 30$\,$M_\odot$. }

\subsection{Three-dimensional Structure and Kinematics}
\label{sec:3Dstructure}
\begin{figure*}
   \centering
   \begin{minipage}{0.48\textwidth}
     \centering
     \includegraphics[width=\linewidth]{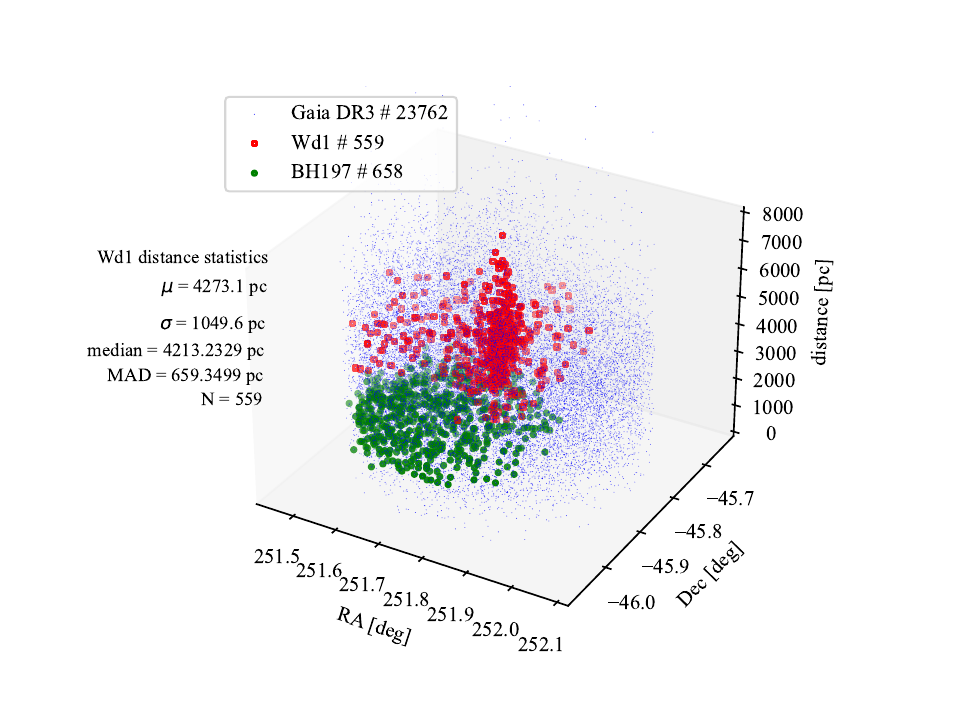}
     \caption{3D spatial distribution (RA, Dec, distance). Wd1 members are shown in red, BH197 in green, and the field population in blue. Distance statistics for Wd1 are reported in the panel.}
     \label{fig:3d_spatial}
   \end{minipage}\hfill
   \begin{minipage}{0.48\textwidth}
     \centering
     \includegraphics[width=\linewidth]{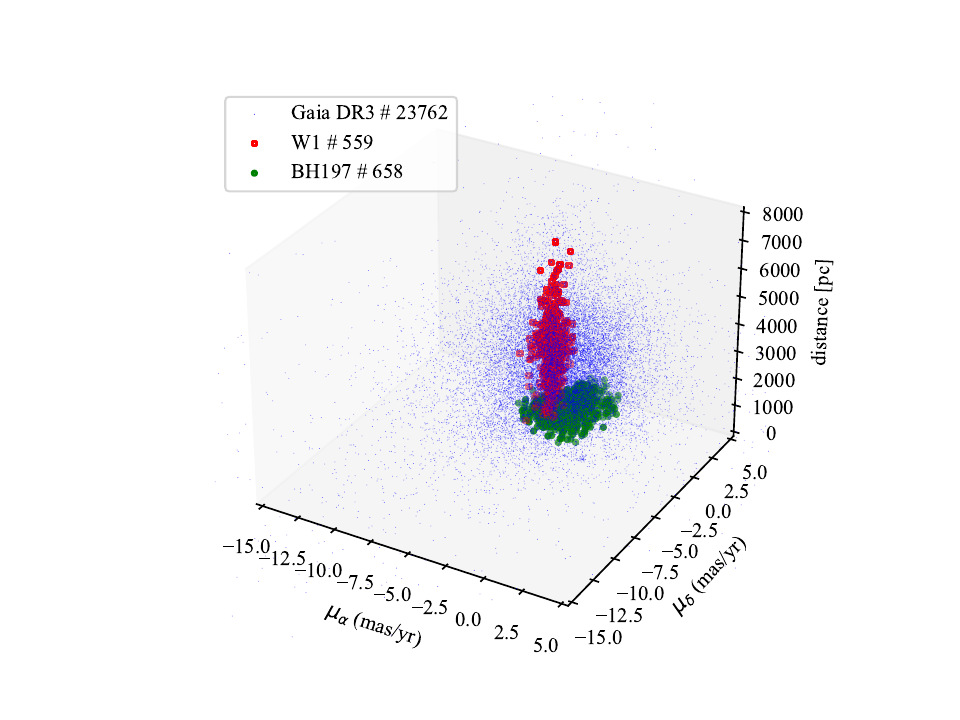}
     \caption{3D kinematic distribution ($\mu_{\alpha}$, $\mu_{\delta}$, distance). The plot highlights the coherence of the proper motion vectors for the cluster members relative to the Galactic background.}
     \label{fig:3d_pm}
   \end{minipage}
\end{figure*}

The final validation of the membership selection was performed by visualizing the identified groups in a 3D parameter space. In particular in the ($\alpha$, $\delta$, d) space and in the ($\mu_\alpha*$, $\mu_\delta$, d) space. 
As illustrated in Fig. \ref{fig:3d_spatial}, we clearly distinguish two physically clustered systems superimposed on the background Galactic population. The main cluster, Wd1 (red squares), shows a high spatial concentration in the RA-Dec plane and is located at a median distance of $d = 4.2$ kpc (with a MAD of $\sim$0.7 kpc), confirming its status as a massive, distant young cluster deeply embedded in the disk of our Galaxy. In contrast, the BH197 cluster (green circles) is resolved as a foreground structure.

The kinematic integrity of these structures is further demonstrated in the 3D proper motion space ($\mu_{\alpha*}, \mu_{\delta}, d$), that we show in Fig.\,\ref{fig:3d_pm}. The members of Westerlund 1 exhibit a remarkably tight distribution in proper motions, clearly offset from the field stars.  The separation in both velocity and distance space between Wd1 and BH197 confirms that the region consists of multiple, independent episodes of star formation located at different depths along the same line of sight.
\section{Results}
\label{sec:kinematics_results}
\subsection{Unbound members}
\label{sec:unbound}
To further characterize the dynamical state of the identified Wd1 candidate members, we compared their internal kinematics with the theoretical escape speed profile of the cluster.
This comparison allows us to identify whether the cluster includes members  that are gravitationally unbound and are 
recently escaped  from the cluster potential well.

As a first step, we transformed the proper motions into internal 2D tangential speeds. We defined the kinematic center of mass using the median proper motions of the sample ($\mu_{\alpha*,\mathrm{bulk}}$, $\mu_{\delta,\mathrm{bulk}}$) and the spatial origin as the median coordinates ($l_{\mathrm{med}}$, $b_{\mathrm{med}}$). 
By subtracting the systemic bulk motion, we shifted to the rest frame of the cluster. The internal 2D speed $v$ for each star was then computed adopting a \rev{fixed} distance of $d = 4.21$\,kpc. To ensure a statistically robust identification of the escaping stars, we performed a rigorous error propagation. 

The total internal \rev{speed} $v$ of each star on the sky plane in the cluster rest frame is defined as $v=\sqrt{v_x^2+v_y^2}$, 
where $x$ and $y$ represent the local orthogonal axis on the sky plane aligned with the directions of Right Ascension ($\alpha$) and Declination ($\delta$), respectively, while $v_x$ and $v_y$ denote the corresponding physical velocity components derived from the relative proper motions
\rev{after bulk-motion subtraction, adopting a fixed cluster distance of $d = 4.21$\,kpc.}

The formal uncertainty $\sigma_v$ on the total internal velocity was calculated for each source by taking into account the correlation of  the uncertainties in right ascension and declination.
The variance of the total velocity $\sigma_v^2$ was computed using the full covariance matrix:
\begin{equation}
\sigma_v^2=\left(\frac{v_x}{v}\right)^2\sigma_{v_x}^2+\left(\frac{v_y}{v}\right)^2\sigma_{v_y}^2+2\left(\frac{v_x}{v}\right)\left(\frac{v_y}{v}\right)\mathrm{cov}(v_x,v_y)
\end{equation}
where $\sigma_{v_x}$ and $\sigma_{v_y}$ are the physical velocity errors converted from the \textit{Gaia} DR3 formal proper motion uncertainties ($\sigma_{\mu_{\alpha*}}$, $\sigma_{\mu_\delta}$), and the covariance term is defined as:
\begin{equation}
\mathrm{cov}(v_x,v_y)=\rho_{x,y}\sigma_{v_x}\sigma_{v_y},
\end{equation}
where $\rho_{x,y}$ corresponds to the \texttt{pmra\_pmdec\_corr} parameter provided in the \textit{Gaia} DR3 archive.

The theoretical escape speed profile, $v_{\mathrm{esc}}(r)$, was estimated under the assumption of a spherically symmetric potential. For a test particle at a  distance $r$ from the cluster center, the escape speed is given by $v_{\mathrm{esc}} = \sqrt{2GM/r}$, where  $M = 5 \times 10^4 M_\odot$ \rev{\citep{ande17}} is the adopted total mass of Wd1. While this point-mass approximation may overestimate the escape speed in the innermost core region, it provides a good estimate of the outer cluster envelope where escaping stars are expected to reside.

To prevent sources with large astrometric uncertainties from being falsely identified as unbound, we applied a stricter selection criterion. A star is classified as unbound only if its measured internal velocity, minus its $1\sigma$ uncertainty, exceeds the local escape speed:
\begin{equation}
    v - 1\sigma_v > v_{\mathrm{esc}}(r)
\end{equation}

\begin{figure}[t!]
   \centering
   \includegraphics[width=\columnwidth]{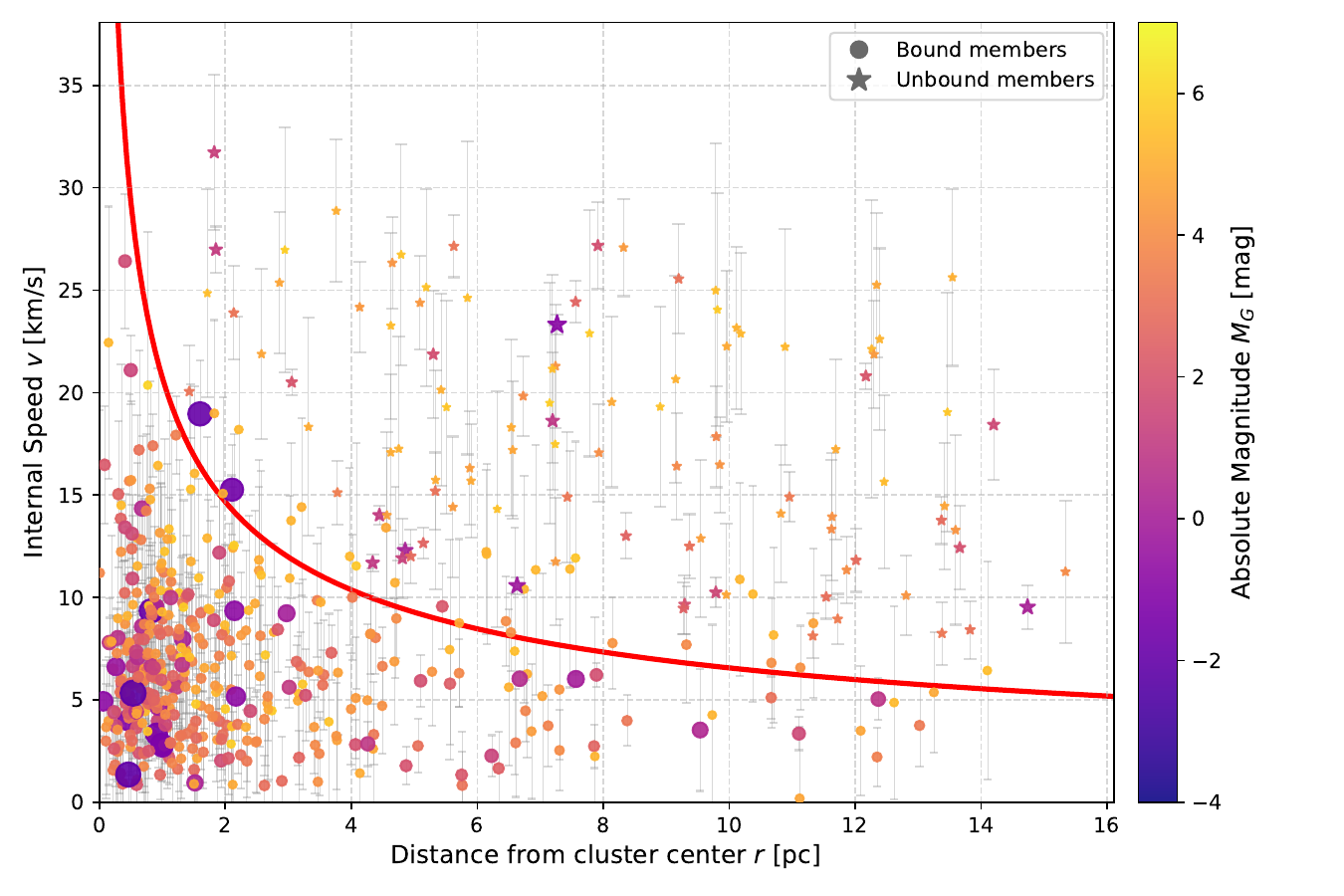}
   \caption{Internal 2D tangential speed of the \textit{Gaia} DR3 members as a function of their projected distance from the cluster center. The solid red line represents the theoretical escape speed profile $v_{\mathrm{esc}} = \sqrt{2GM/r}$, assuming a total cluster mass of $5 \times 10^4 M_\odot$. Gravitationally bound stars ($v + 1\sigma \le v_{\mathrm{esc}}$) are denoted by circles, whereas robustly unbound members ($v - 1\sigma > v_{\mathrm{esc}}$) are marked by star symbols. The markers are color-coded according to the absolute magnitude $M_G$ of each star. Furthermore, the symbol sizes $s$ are scaled to reflect the intrinsic stellar luminosity as $s = 10 + 50 \times 10^{-0.2 M_G}$.}
   \label{fig:vel_fuga}
\end{figure}

Figure~\ref{fig:vel_fuga} illustrates the distribution of the internal tangential \rev{speed} as a function of the projected distance from the cluster center. With  our criterion we  identify 103 unbound members that are highlighted in Fig.~\ref{fig:vel_fuga} with star symbols. 
We note that for simplicity of nomenclature, the bound sample is defined here as the remaining co-moving population after excluding the  unbound members. Consequently, this sample may inherently include a fraction of stars with velocities close to or marginally above the escape velocity floor.

\rev{For the unbound sample, the observed velocity dispersions along Galactic longitude ($v_l$) and latitude ($b$), evaluated in the cluster rest frame after bulk-motion subtraction, yield $\sigma_{\mathrm{obs}}(v_l) = 14.93~\mathrm{km\,s^{-1}}$ and $\sigma_{\mathrm{obs}}(v_b) = 11.22~\mathrm{km\,s^{-1}}$ (with an anisotropy ratio $\sigma_{\mathrm{obs}}(v_b)/\sigma_{\mathrm{obs}}(v_l) \approx 0.75$). 
Analogously, for the bound sample, the observed velocity dispersions are $\sigma_{\mathrm{obs}}(v_l) = 5.83~\mathrm{km\,s^{-1}}$ and $\sigma_{\mathrm{obs}}(v_b) = 5.40~\mathrm{km\,s^{-1}}$, 
\rev{consistent with an isotropic velocity dispersion in the cluster core}
$\sigma_{\mathrm{obs}}(v_b)/\sigma_{\mathrm{obs}}(v_l) \approx 0.93$.}

We  cross-matched the bound and unbound populations with the X-ray catalogue to investigate the 
possible differences in their high-energy emission properties.
Among the 421 bound stars, 273 objects are detected in X-rays, while 148 do not 
show any X-ray counterpart, corresponding to an X-ray detection fraction of 
$\sim65\%$. In contrast,  25 of the 103 unbound stars are associated with 
X-ray emission, yielding a detection fraction of $\sim24\%$. Therefore, the 
X-ray detected population is significantly more concentrated among the 
kinematically bound stars, suggesting that the bound population contains a 
larger fraction of X-ray emitters.

While the X-ray detection fraction of the unbound population is lower than that of the dense core, a 24\% detection rate is still significantly higher than the 13\% fraction observed among the discarded field stars. This provides additional evidence that the unbound sample is genuinely associated with Wd1.

Among the 103  unbound sources, we isolate the massive candidate W238 (\texttt{source\_id} 5940105904021494144), a spectroscopically confirmed B1~Iab luminous supergiant \citep{cott12,clar20}, which matches our astrometric position within a tight separation of just $0.29^{\prime\prime}$. 

Furthermore, out of the 359 cluster members in our sample selected by adopting the \citet{negu22} spatial and color constraints *$r \le 3.5^\prime$ and $BP-RP > 4.0$ mag), 346 are classified as gravitationally bound and 13 as unbound. Among the 303 cross-matched with the \citet{negu22} catalogue, we find 297 bound and 6 unbound members within our kinematic census.

\subsection{Kinematic Modeling of Westerlund\,1}
\label{sec:kinematics}

Besides unbound members, we
find evidence of velocity spreads in excess of measurement errors.
To investigate the internal macroscopic dynamics of Wd1, we performed a 2D kinematic analysis of the proper motion fields. To minimize the impact of astrometric noise and avoid spurious signals from low-quality measurements, we applied a strict quality cut across the master catalog, retaining only those stars with a total proper motion uncertainty smaller than $0.5\,\mathrm{mas}\,\mathrm{yr}^{-1}$ (corresponding to $\approx 10\,\mathrm{km\,s^{-1}}$ at the adopted cluster distance).

\subsubsection{The Kinematic Model and error treatment}
\label{subsubsec:model_propagation}

The simplest non-rigid model of internal kinematics is a
first-order 2D Taylor expansion of the velocity field. The observed proper motions in the Galactic coordinate system ($\mu_{l*}, \mu_b$) were expressed as a linear function of the position relative to a fixed spatial origin ($l_0, b_0$):
\begin{equation}
    v_l = v_l^0 + \frac{\partial v_l}{\partial l}\delta l + \frac{\partial v_l}{\partial b}\delta b ,
\end{equation}
\begin{equation}
    v_b = v_b^0 + \frac{\partial v_b}{\partial l}\delta l + \frac{\partial v_b}{\partial b}\delta b ,
\end{equation}
where $\delta l = (l - l_0)\cos(b_0)$ and $\delta b = b - b_0$ represent the  spatial projections on the celestial sphere and  $v_l^0$ and $v_b^0$ denote the bulk motion components of the system.
The four spatial partial derivatives define the 2D velocity 
gradient tensor (or Jacobian matrix of the velocity field), denoted as:
\begin{equation}
J = \begin{pmatrix} 
\frac{\partial v_l}{\partial l} & \frac{\partial v_l}{\partial b} \\ 
\frac{\partial v_b}{\partial l} & \frac{\partial v_b}{\partial b} 
\end{pmatrix} .
\end{equation}

A critical aspect of this modeling is the rigorous treatment of individual astrometric uncertainties. 
We fully accounted for the correlated nature of the \textit{Gaia}~DR3 astrometry by rigorously propagating the full error covariance matrix from the system in which the proper motions are released, i.e. 
the Equatorial frame (ICRS),  to the Galactic frame. For each star, we extracted the formal standard errors ($\sigma_{\mu_{\alpha*}}, \sigma_{\mu_\delta}$) and the correlation coefficient $\rho(\mu_{\alpha*}, \mu_\delta)$ directly from the \textit{Gaia} catalog. Using the localized projection Jacobian matrix ($R$), the transformation of the ICRS covariance matrix ($C_{\mathrm{ICRS}}$) onto the Galactic plane ($C_{\mathrm{gal}} = R \, C_{\mathrm{ICRS}} \, R^T$) yields the mathematically exact individual uncertainties along the Galactic longitude and latitude axis:
\begin{equation}
    \sigma_{v_l}^2 = R_{11}^2 \sigma_{\mu_{\alpha*}}^2 + 2 R_{11} R_{12} \, \rho(\mu_{\alpha*}, \mu_\delta) \sigma_{\mu_{\alpha*}} \sigma_{\mu_\delta} + R_{12}^2 \sigma_{\mu_\delta}^2 ,
\end{equation}
\begin{equation}
    \sigma_{v_b}^2 = R_{21}^2 \sigma_{\mu_{\alpha*}}^2 + 2 R_{21} R_{22} \, \rho(\mu_{\alpha*}, \mu_\delta) \sigma_{\mu_{\alpha*}} \sigma_{\mu_\delta} + R_{22}^2 \sigma_{\mu_\delta}^2,
\end{equation}
where $R_{ij}$ (with $i, j = 1, 2$) represent the individual elements of the localized 
Jacobian matrix $R$, computed  for each source.

To fit the 2D kinematic parameters we used 
the \texttt{scipy.optimize.curve\_fit} Python package. We evaluated both a standard ordinary least-squares (OLS) fit and a weighted least-squares (WLS) fit, using the above errors. We found that the unweighted OLS approach yields significantly more stable and physically consistent results across the different sub-samples we evaluated. 
\rev{This is because WLS is heavily dominated by a small handful of bright stars with very small formal uncertainties ($1/\sigma^2$), which artificially reduces the spatial coverage and makes the derived gradients overly sensitive to local velocity fluctuations. This effect is directly driven }
 by the distribution of the formal astrometric uncertainties characterizing the Gaia measurements \citep{fabr21}. An analysis of the proper motion error distributions reveals that while the bulk of the cluster members have uncertainties in the range of $\sim 0.05-0.35$ mas/yr, the distribution exhibits a secondary peak extending towards extremely small formal errors (down to $\sim 0.02-0.03$ mas/yr), especially for the brightest targets.

In a standard WLS fit, individual stars are weighted by the inverse variance of their measurements ($w_i \propto 1/\sigma_i^2$). Due to this quadratic dependence, even a small subset of stars with very low formal Gaia errors receives an overwhelming  statistical weight, effectively anchoring the kinematic plane to a handful of sources. 

The observed proper motion of any cluster member is the combination of the bulk systemic kinematics and the star's own intrinsic orbital motion (i.e., the intrinsic velocity dispersion of the cluster). For the brightest stars with the lowest instrumental errors, this intrinsic velocity dispersion easily dominates over the formal astrometric precision. Consequently, weighting strictly by the Gaia instrumental errors forces the model to over-fit the peculiar orbital motions of a few bright stars, distorting  the macroscopic kinematic gradients of the system. 

By employing an unweighted fit, we treat the stellar population homogeneously.  This ensures that the derived linear gradients accurately map the macroscopic structural kinematics of the cluster, such as its expansion and rotation, rather than being biased by the local
kinematics of the highest-precision targets.

Besides the bulk motion, the fit allows us to derive the following physical quantities: the apparent expansion rate (Divergence, defined as the trace of the
velocity gradient tensor, $\partial v_l / \partial l + \partial v_b / \partial b$) and the rotation/shear rate (Vorticity, $\partial v_b / \partial l - \partial v_l / \partial b$). Their associated uncertainties are obtained by propagating in quadrature the standard errors of the independent linear fits.

Furthermore, by constructing the symmetric deformation tensor $S = \frac{1}{2}(J + J^T)$ from the Jacobian matrix $J$, we extracted the Position Angle ($PA_{\mathrm{gal}}$, measured from Galactic North toward East) of the principal axis of maximum expansion.

We computed the reduced Chi-squared ($\chi^2_\nu$) independently for both Galactic velocity components. For a given sample of $N_{\star}$ stars, the $\chi^2_\nu$ estimator along the coordinate direction $k \in \{l, b\}$ is defined as:

\begin{equation}
    \chi^2_{\nu, k} = \frac{1}{N_{\star} - 3} \sum_{i=1}^{N_{\star}} \left( \frac{v_{k, i} - v_{\rm model, \textit{k}}(x_i, y_i)}{\sigma_{v_k, i}} \right)^2
\end{equation}

where $v_{k, i}$ is the observed relative proper motion of the $i$-th star, $v_{\rm model, \textit{k}}(x_i, y_i)$ is the velocity predicted by the best-fit linear regression plane evaluated at the relative coordinates $(x_i, y_i)$, and $\sigma_{v_k, i}$ is the individual formal uncertainty propagated from the \textit{Gaia} DR3 catalog. 

The denominator $N_{\star} - 3$ represents the degrees of freedom ($\mathrm{dof}$), accounting for the three free parameters of the first-order Taylor expansion model fitted to each spatial dimension. These reduced $\chi^2_\nu$ values serve as a proxy for the internal kinematic scatter of each sub-sample relative to the formal astrometric errors.

\subsubsection{Multi-sample Kinematics}
\label{subsubsec:multi_pop}
The identification of many escaped stars 
(Sect.\,\ref{sec:unbound})
suggests that the internal dynamical structure of the cluster may be quite complex. 
To map this spatial and kinematic complexity, we fixed the coordinate center ($l_0, b_0$) to the geometric barycenter of the gravitationally bound core ($l_0 = 339.549^\circ$, $b_0 = -0.401^\circ$). We then 
repeated the fit across five physically
distinct stellar subsamples.
The results of this kinematic tomography, converted into physical units assuming a \rev{fixed} cluster distance of $d = 4.21\,\mathrm{kpc}$, are summarized in Table~\ref{tab:kinematics_all_samples}. 

\input{FIGURES/taylor_exp_fit_all_samples_spatial_errpa_noerr}


Besides the complete sample of cluster members ('DBSCAN All'), we separately fitted the 'Bound' and 'Unbound' sub-samples to investigate discrepancies in their kinematics. 
The fit was also performed on the 'Field' sample. Furthermore, to test whether the kinematic behavior of the Unbound stars mimics that of the field population rather than the cluster members,
as expected if the Unbound sample were dominated by non-member contaminants,
we built a composite sample by adding 100 field stars, randomly extracted within the cluster's spatial region, to the Bound sample. The final parameters and uncertainties for this test case report the median values and median formal errors obtained from 1000 Monte Carlo iterations.

\subsubsection{Fitting results of the main sample}
\label{subsubsec:fitresults}
Regarding the entire sample of cluster members (DBSCAN All sample),
we obtained the following results.
We derived a bulk motion of \rev{$v_l^0=-85.07\pm0.38\,$\,km/s and 
$v_b^0=-13.74\pm0.30\,$\,km/s, respectively. }
The other motion components derived from our fit are the  expansion and the vorticity. 
\begin{figure}[t]
\centering
\includegraphics[width=\columnwidth]{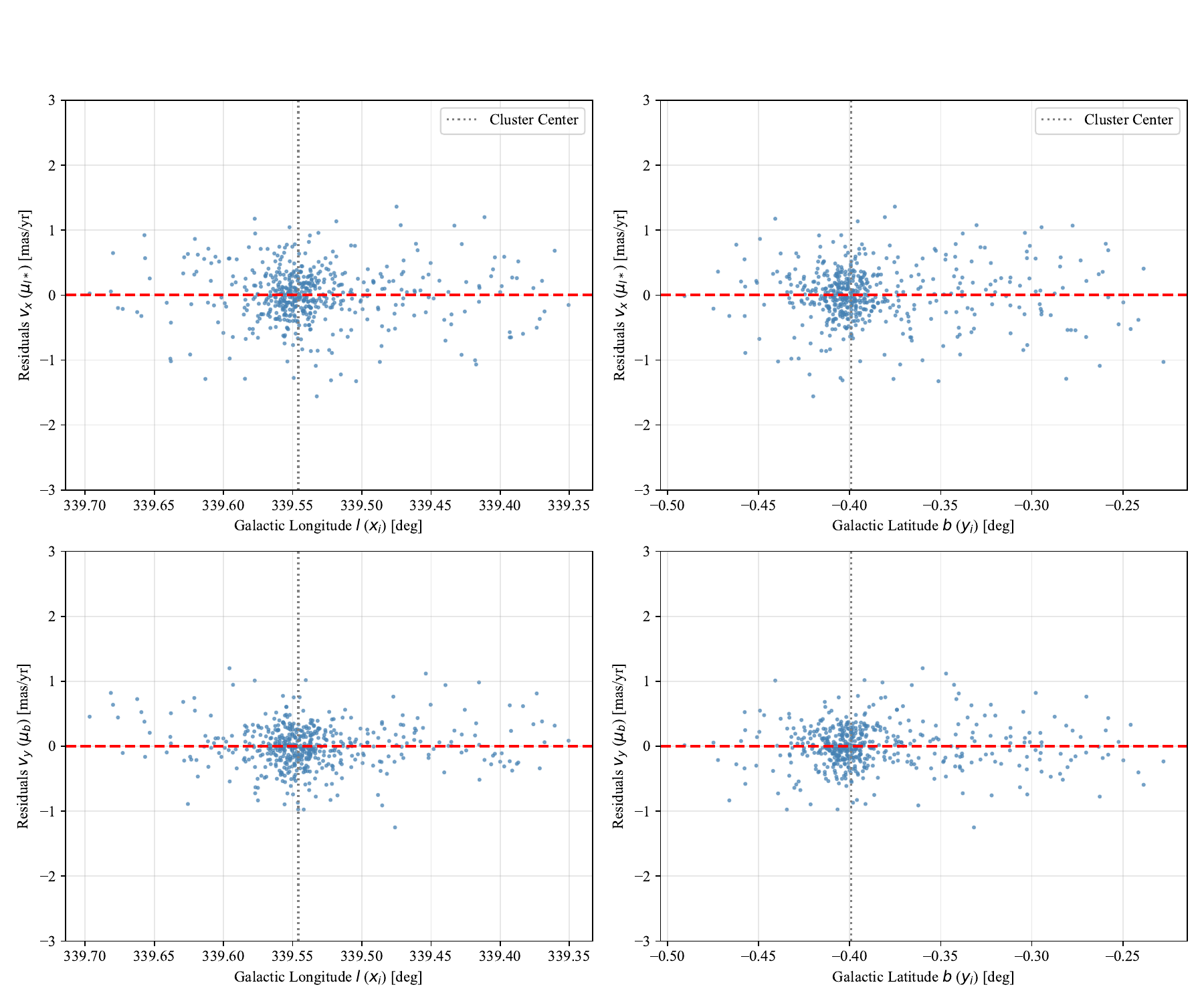}
\caption{Residuals of the kinematic fit ($\mu_{l*}$ and $\mu_b$) vs. absolute Galactic coordinates. The dashed red line marks zero. }
\label{fig:residuals}
\end{figure}
\begin{figure}[h!]
\centering
\includegraphics[width=\columnwidth]{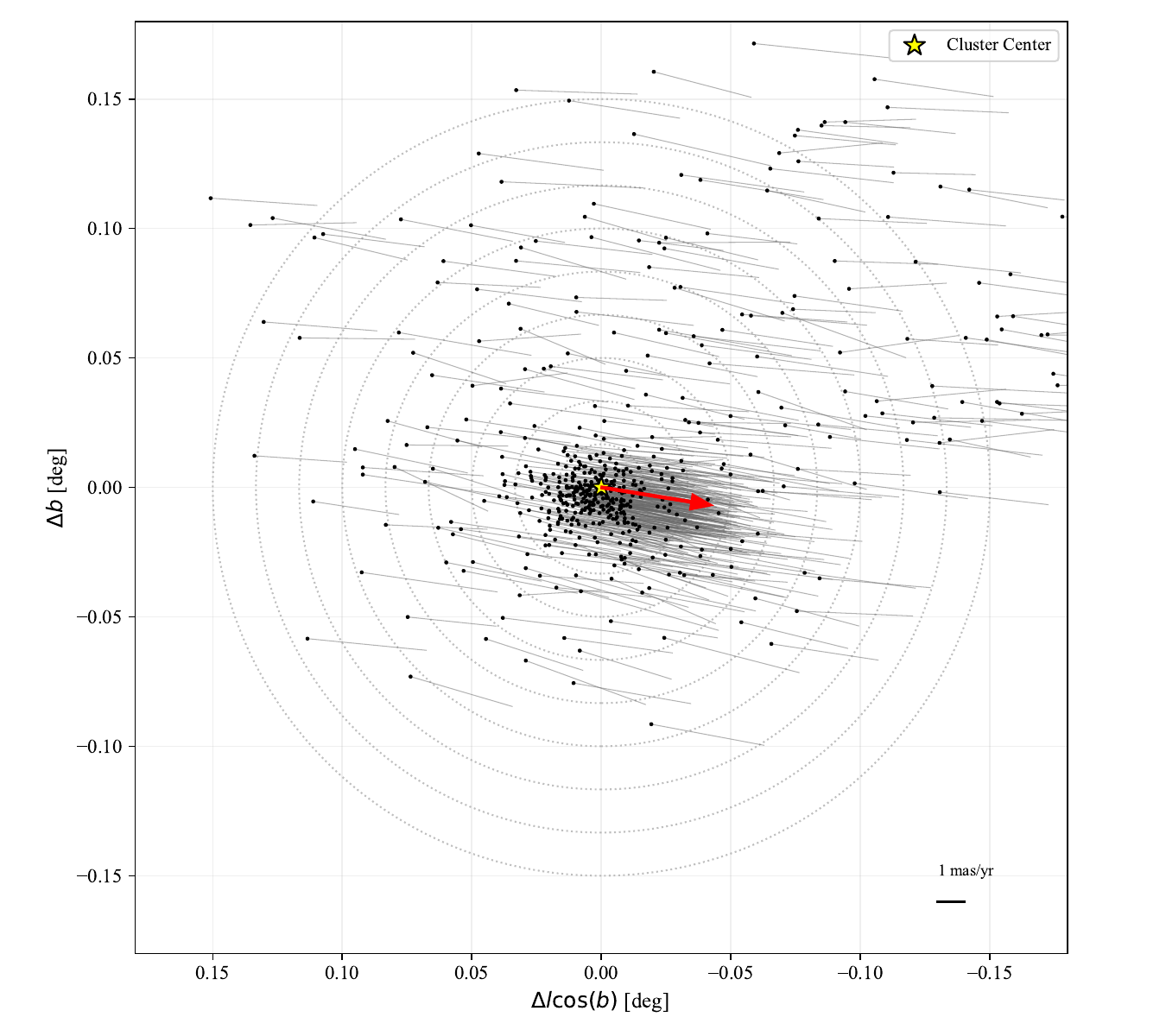}
\caption{Absolute proper motion vector field of the selected Wd1 members, mapped in spatial coordinates relative to the cluster center ($\Delta l \cos b$, $\Delta b$). The grey thin segments indicate the individual stellar velocities, while the thick red arrow originating from the origin (yellow star) represents the mean systemic bulk motion derived from our kinematic fit.}
\label{fig:absolute_kinematics}
\end{figure}

The  expansion or 
    divergence  is $2.13 \pm 0.52$ mas\,yr$^{-1}$\,deg$^{-1}$ or
    $0.58 \pm 0.14$ km\,s$^{-1}$\,pc$^{-1}$,
    calculated as $\frac{\partial v_l}{\partial l} + \frac{\partial v_b}{\partial b}$, the trace of the velocity gradient tensor.
     Our results indicate that
    the divergence is  dominated by the vertical gradient ($\partial v_b / \partial b = 0.47 \pm 0.10$ km\,s$^{-1}$\,pc$^{-1}$). This indicates an out-of-plane expansion.
    
 The vorticity and shear  amounts to $-2.22 \pm 0.54$ mas/yr/deg or
$-0.60 \pm 0.15$ km\,s$^{-1}$\,pc$^{-1}$,
 calculated as $\frac{\partial v_b}{\partial l} - \frac{\partial v_l}{\partial b}$. For the Gaia DR3 Wd1 members, the vorticity is dominated by a strong cross-gradient $\partial v_b / \partial l = -0.41 \pm 0.08$\,km\,s$^{-1}$\,pc$^{-1}$. \del{This signal ($>5\sigma$) is likely a  tracer of the tidal torque exerted by the Galactic potential.} \rev{The spatial kinematic gradients derived from our analysis can originate from a variety of internal dynamical processes, such as  internal rotation, or shear; nevertheless, external influences such as Galactic tidal torques may also contribute to the observed velocity field.}

To check for systematic trends,
we analyzed the residuals, which are shown in Fig. \ref{fig:residuals}. These do not show trends since they are randomly distributed around zero, validating the adopted linear model. 
The high dispersion observed in the residuals ($\sigma_{\rm res, \mathit{l}} = 0.421$\,mas\,yr$^{-1}$ and $\sigma_{\rm res, \mathit{b}} = 0.331e$\,mas\,yr$^{-1}$) reflects the intrinsic velocity dispersion of Wd1, which significantly exceeds Gaia's typical instrumental noise for this sample, $\sigma_{\rm Gaia, \mathit{l}} = 0.115$\,mas\,yr$^{-1}$ and $\sigma_{\rm Gaia, \mathit{b}} = 0.145$\,mas\,yr$^{-1}$. These latter uncertainties  were derived by propagating the full Gaia DR3 covariance matrix (including the correlation coefficient $\rho(\mu_\alpha^*, \mu_\delta)$) through the local ICRS-to-Galactic coordinate transformation matrix.

The reduced chi-square values ($\chi^2_{\nu, l} \approx 26.2$ and $\chi^2_{\nu, b} \approx 15.7$) are significantly higher than unity. This excess cannot be attributed to Gaia errors alone but likely represents the detection of the intrinsic velocity dispersion of the cluster.

To isolate the intrinsic 1D velocity dispersion of the cluster ($\sigma_{\mathrm{int}}$) from the observed residuals spread ($\sigma_{\mathrm{res}}$), we performed a quadrature subtraction of the mean formal Gaia DR3 uncertainties ($\sigma_{\mathrm{Gaia}}$):
\begin{equation}
\sigma_{\mathrm{int}} = \sqrt{\sigma_{\mathrm{res}}^2 - \sigma_{\mathrm{Gaia}}^2} .
\end{equation} 
This quadrature subtraction was executed independently for the Bound and Unbound samples in order to evaluate
the dynamical state of the two samples.

For the Bound core sample, the empirical standard deviations of the relative velocity distributions are \rev{$\sigma_{\mathrm{obs},l} = 5.83~\mathrm{km\,s^{-1}}$ and $\sigma_{\mathrm{obs},b} = 5.40~\mathrm{km\,s^{-1}}$} (see Sect.\,\ref{sec:unbound}). Considering the median Gaia instrumental noise derived for this sub-sample 
\rev{($\sigma_{\mathrm{Gaia},l} = 2.22~\mathrm{km\,s^{-1}}$ and $\sigma_{\mathrm{Gaia},b} = 2.84~\mathrm{km\,s^{-1}}$)} yields intrinsic velocity dispersions of:
\rev{\begin{equation}\sigma_{\mathrm{int},l} = \sqrt{5.83^2 - 2.22^2} = 5.39~\mathrm{km\,s^{-1}} ,\end{equation}\begin{equation}\sigma_{\mathrm{int},b} = \sqrt{5.40^2 - 2.84^2} = 4.59~\mathrm{km\,s^{-1}} ,\end{equation}}corresponding to an intrinsic anisotropy ratio of $\sigma_{\mathrm{int},b}/\sigma_{\mathrm{int},l} = 0.85$.

For the Unbound sample, the empirical observed dispersions are \rev{$\sigma_{\mathrm{obs},l} = 14.93~\mathrm{km\,s^{-1}}$ and $\sigma_{\mathrm{obs},b} = 11.22~\mathrm{km\,s^{-1}}$} (see Sect.\,\ref{sec:unbound}). Applying the same quadrature subtraction with the corresponding sub-sample instrumental noise \rev{($\sigma_{\mathrm{Gaia},l} = 2.81~\mathrm{km\,s^{-1}}$ and $\sigma_{\mathrm{Gaia},b} = 3.15~\mathrm{km\,s^{-1}}$) }results in:\rev{\begin{equation}\sigma_{\mathrm{int},l} = \sqrt{14.93^2 - 2.81^2} = 14.66~\mathrm{km\,s^{-1}} ,\end{equation}\begin{equation}\sigma_{\mathrm{int},b} = \sqrt{11.22^2 - 3.15^2} = 10.77~\mathrm{km\,s^{-1}} ,\end{equation}}yielding an intrinsic anisotropy ratio of $\sigma_{\mathrm{int},b}/\sigma_{\mathrm{int},l} = 0.73$.

To visually show the overall kinematics of the Wd1 cluster, we first present the 2D absolute proper motion vector field in Fig.~\ref{fig:absolute_kinematics}. This raw astrometric map is heavily dominated by the systemic bulk motion of the cluster ($\mu_{l*} \approx -4.26$ mas\,yr$^{-1}$, $\mu_b \approx -0.70$ mas\,yr$^{-1}$), causing the vast majority of the stellar motions to point coherently toward the Galactic South-West, as indicated by the red arrow. 
\begin{figure}[htbp]
\centering
\includegraphics[width=\columnwidth]{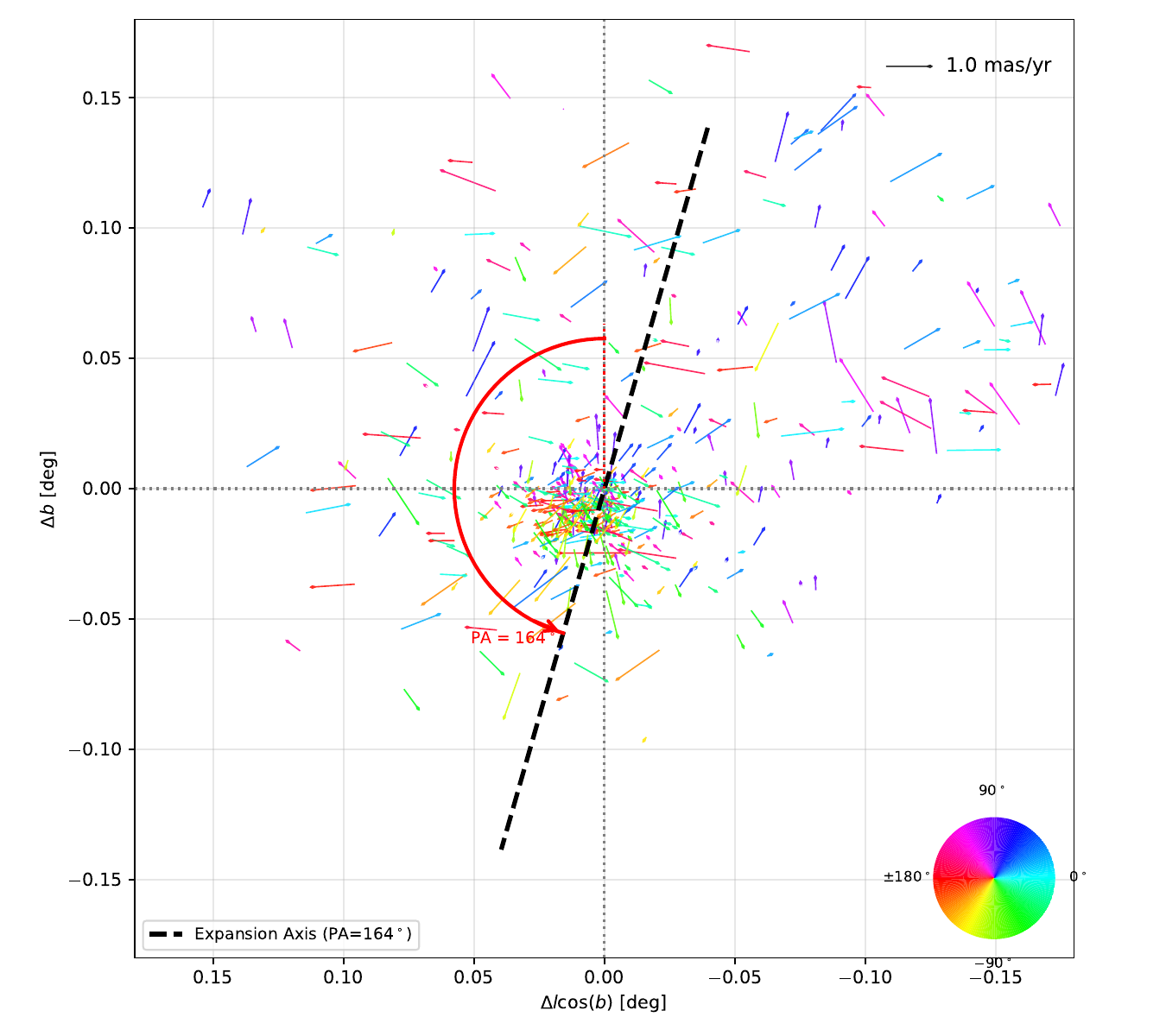}
\vspace{0.2cm}
\includegraphics[width=0.96\columnwidth]{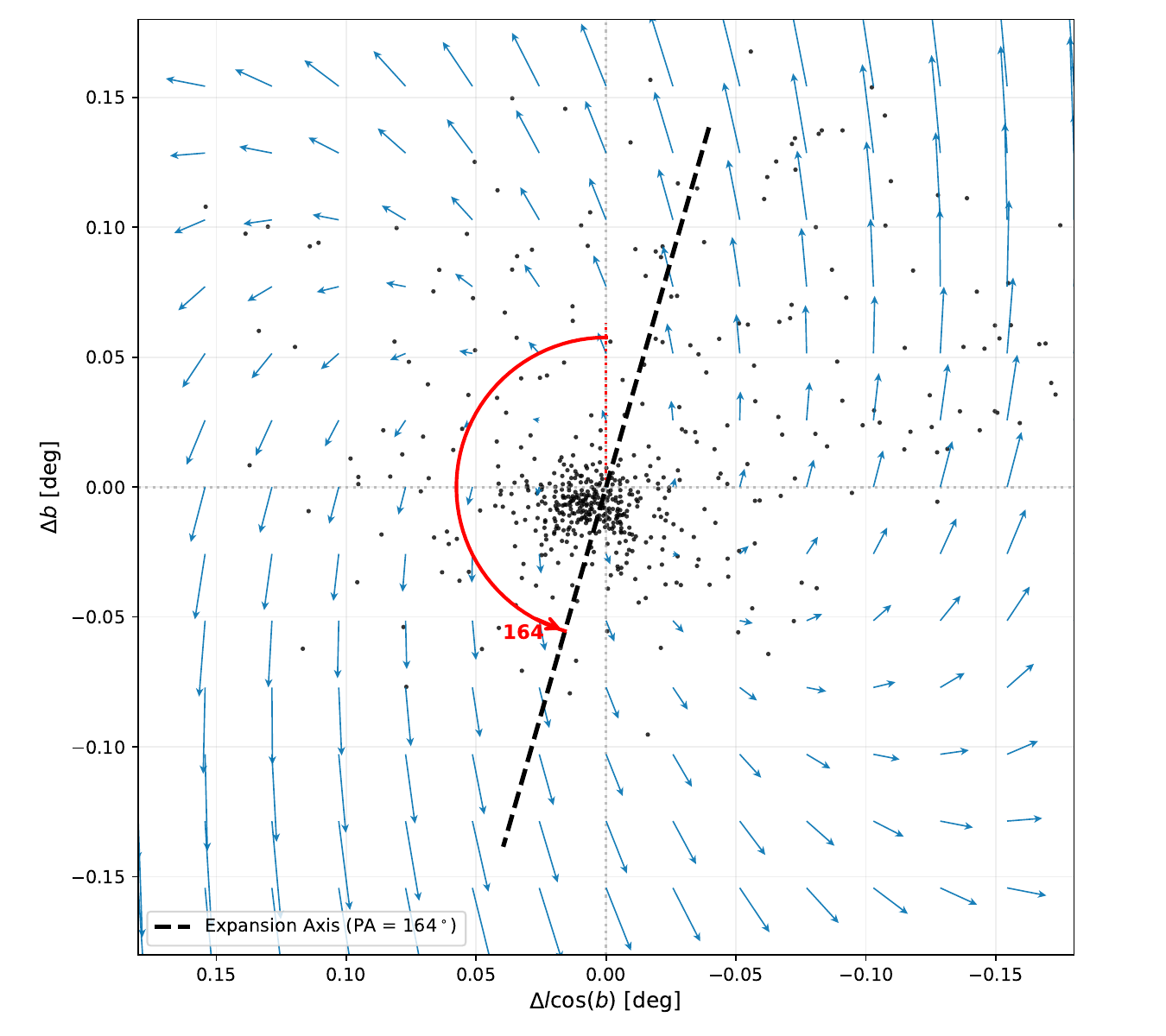}
\caption{Internal kinematics of the Westerlund 1 cluster. \textit{Top:} Internal proper motion field obtained by subtracting the median systemic proper motion from the absolute motions of the stars. Proper motion vectors are colour-coded by the position angle of their motion, as per the colour wheel in the lower right-hand corner. \textit{Bottom:} Average macroscopic velocity field derived from the first-order Taylor expansion, evaluated on a regular grid. The actual spatial positions of the identified members are overlaid as reference points. In both panels, the dashed line marks the principal kinematic expansion axis, analytically determined from the fitted velocity gradient tensor.}
\label{fig:internal_kinematics}
\end{figure}

To reveal the underlying internal dynamics, we subtracted this systemic translation from the absolute motion of each individual star. The resulting relative proper motions are presented in the top panel of Fig.~\ref{fig:internal_kinematics}. This reference frame, centered on the proper motion, shows the internal expansion and the rotation. The observed vector field appears quite chaotic, since it
reflects the significant intrinsic velocity dispersion characterizing the massive stars in the Wd1 core.

The anisotropic expansion direction corresponds to a position angle (PA) of $164^\circ$ (dashed line), where the PA is conventionally defined in the Galactic coordinate system as the angle measured counterclockwise (East) starting from Galactic North (i.e., towards positive Galactic latitude $+b$, corresponding to $0^\circ$, and increasing towards positive Galactic longitude $+l$). 
This PA was analytically determined by extracting the dominant eigenvector from the velocity gradient tensor derived from our fit.
The global apparent expansion is primarily aligned along the North-West to South-East direction. This expansion orientation is in  agreement with the  PA$_{\rm gal}$ of $133.6^\circ \pm 8.0^\circ$ derived by \citet{orde26b} using independent VVVX infrared data. This marginal consistency  between the two completely independent estimates can be explained by considering that the two datasets probe  different stellar populations. While our \textit{Gaia} DR3 analysis maps the less obscured members and the outer regions of the cluster footprint, the VVVX data successfully scan the highly obscured, deeply embedded core. The fact that both datasets reveal a similar macroscopic kinematic axis demonstrates that this asymmetric expansion is a global, cluster-wide structural phenomenon.

To provide a clear physical representation of the macroscopic gradients extracted by our linear fit and disentangle them from the intrinsic velocity dispersion, the bottom panel of Fig.~\ref{fig:internal_kinematics} displays the model velocity field computed from the parameters given in Table~\ref{tab:kinematics_all_samples} for the DBSCAN All sample and evaluated on a regular grid. This average macroscopic model show the curvature which is the physical visualization of the rotation (vorticity, $\sim -0.60$ km s$^{-1}$ pc$^{-1}$) coupled with the cluster's expansion.

\subsubsection{Comparative analysis across stellar sub-samples and control tests}

To assess the robustness of the detected kinematic patterns, we compare the results obtained from the full DBSCAN All sample with those derived from the independent Bound and Unbound sub-samples (see Table~\ref{tab:kinematics_all_samples}). For both the bound core and the escaping stellar populations, we resolve spatial velocity gradients that are  consistent with the global expansion and rotation trends found for the entire cluster. Specifically, the divergence and vorticity parameters remain positive and negative, respectively, across all three samples. However, due to the smaller number of sources available in the individual sub-samples, these kinematic signatures are not significant for the bound sample and are detected at only $\sim 2-3\sigma$  statistical significance with respect to the robust $\sim 4-5\sigma$ threshold achieved by the complete member catalog. 

In  contrast, the analysis of the control samples reinforces the physical reality of the cluster's internal kinematics and member selection. The regression planes fitted to the pure \texttt{Field} sample yield spatial gradients and structural parameters (Divergence and Vorticity) that are statistically consistent with zero within their large formal uncertainties. Similarly, the synthetic control test consisting of the cluster core contaminated with randomly extracted field stars (\texttt{Bound + Field (MC 1000x)}) shows a significant dilution of the kinematic signals. In this mixed configuration, the macroscopic divergence drops to a negligible value ($\rev{0.07} \pm 0.31~\mathrm{km\,s^{-1}\,pc^{-1}}$), and the coherent expansion axis becomes poorly constrained. 

These control tests \del{conclusively} confirm that the systematic divergence and vorticity mapped in the DBSCAN All sample are genuine, co-moving dynamical properties of the Westerlund~1 cluster, rather than artificial projections or geometric biases induced by background Galactic field stars passing through the cluster footprint with similar apparent motion.
\section{Discussion}
\label{sec:discussion}

\subsection{Cluster and Galactic Potential}
\label{sec:potential}
The evidence of a highly anisotropic, out-of-Galactic plane expansion in Wd1 adds a crucial piece of evidence to the emerging picture of young cluster evolution. Recent 3D kinematic  surveys indicate that macroscopic expansion is a widespread phenomenon \citep{,wrig20}; for instance, \citet{wrig24} demonstrated that up to 94\% of young clusters exhibit active expansion, with 88\% of them expanding asymmetrically. The pronounced anisotropy we resolve in Wd1 closely mirrors specific case studies in the literature, such as the highly anisotropic expansion field detected in the young cluster NGC\,6530 \citep{wrig19}. This strong statistical and morphological convergence implies that the kinematic perturbations and the subsequent anisotropic dissolution we map in Wd1 are not isolated anomalies, but are likely representative of a common evolutionary mechanism operating across many young Galactic systems.

To further investigate how the observed spatial distribution and the  internal kinematics of Wd1 are influenced by the MW, we correlated the cluster morphology and the surrounding interstellar medium with the local effective Galactic potential. 

\begin{figure}[t!]
   \centering
\includegraphics[width=\columnwidth]{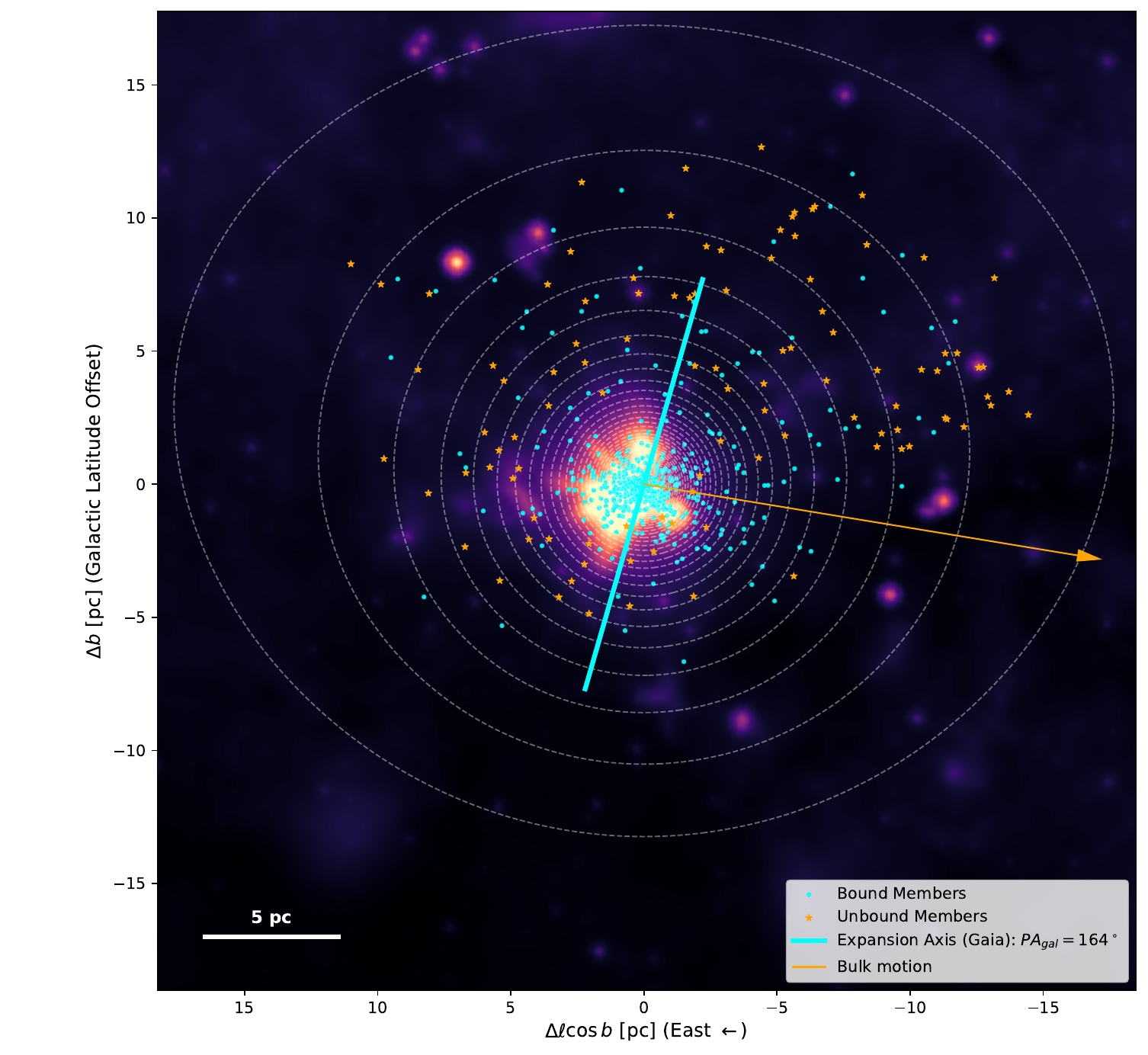}
\caption{Spatial distribution of the identified Wd1 members overlaid on a WISE 22~$\mu$m mid-infrared image. Cyan dots represent gravitationally bound members, while orange stars indicate  unbound members. White contours represent the equipotential curves of the total effective  potential, including the Galactic disk and halo, the cluster's self-gravity, and the centrifugal term. The map is centered on the median coordinates of the identified members. The solid cyan  line represent the principal kinematic expansion axes derived from this work . The orange arrow indicates the direction and magnitude of the mean cluster bulk motion. }
   \label{fig:wise_overlay}
\end{figure}

Fig.~\ref{fig:wise_overlay} shows the identified Gaia DR3 members overplotted on a mid-infrared image from the WISE survey at 22~$\mu$m. In this band, the extended emission is largely dominated by thermal radiation from dust grains heated by the intense ultraviolet radiation field of the cluster's massive stars. The spatial concentration of our selected members strongly corresponds with the peak of this mid-infrared emission, providing an independent morphological validation of our astrometric membership selection.

To model the environmental effects of the MW on Wd1, we calculated the total effective gravitational potential, $\Phi_{\mathrm{tot}}$, which is the sum of the Galactic potential, the centrifugal potential due to the rotating reference frame of the Wd1 orbit around the Galactic center, and the cluster's own self-gravity.

The MW's background potential was modeled using an axisymmetric mass distribution consisting of a stellar disk and a halo. The Galactic disk is represented by a Miyamoto-Nagai potential \citep{miya75}:
\begin{equation}
    \Phi_{\mathrm{disk}}(R, Z) = -\frac{G M_{\mathrm{d}}}{\sqrt{R^2 + \left(a + \sqrt{Z^2 + b^2}\right)^2}}
\end{equation}
where $R$ and $Z$ are the Galactocentric cylindrical coordinates, $G=4.499\times 10^{-3}\,pc^3\,Myr^{-2}\,M_\odot^{-1}$ is the gravitational constant, $M_{\mathrm{d}} = 6.0 \times 10^{10}\,M_\odot$ is the disk mass, and $a = 3500$\,pc and $b = 300$\,pc are the scale length and scale height, respectively. 
For the halo, we adopted a Navarro-Frenk-White (NFW) profile \citep{nava97}:
\begin{equation}
    \Phi_{\mathrm{halo}}(r) = - 4 \pi G \rho_s R_s^3 \frac{\ln(1 + r/R_s)}{r}
\end{equation}
where $r = \sqrt{R^2 + Z^2}$ is the spherical Galactocentric radius, $\rho_s = 0.0085\,M_\odot\,\mathrm{pc}^{-3}$ is the characteristic density, and $R_s = 16\,000$\,pc is the scale radius.

Since the cluster follows a Galactic orbit, we evaluated the potential in a reference frame corotating with Wd1. We introduced a centrifugal potential term:
\begin{equation}
    \Phi_{\mathrm{centr}}(R) = -\frac{1}{2} \omega^2 R^2
\end{equation}
with $\omega \approx 0.0464\,\mathrm{Myr}^{-1}$ being the orbital angular velocity 
for a circular orbit at a Galactocentric distance of $R = 4.2$\,kpc, according to 
the adopted Galactic potential.

Finally, the internal self-gravity of Wd1 was modeled as a Plummer potential \citep{plum11}:
\begin{equation}
    \Phi_{\mathrm{clust}}(d) = -\frac{G M_{\mathrm{c}}}{\sqrt{d^2 + r_{\mathrm{c}}^2}}
\end{equation}
where $d$ is the distance from the cluster center, $M_{\mathrm{c}} = 5 \times 10^4\,M_\odot$ is the estimated total mass of the cluster, and $r_{\mathrm{c}} = 1.0$\,pc is the core radius.

The total effective potential is thus given by:
\begin{equation}
    \Phi_{\mathrm{tot}} = \Phi_{\mathrm{disk}} + \Phi_{\mathrm{halo}} + \Phi_{\mathrm{centr}} + \Phi_{\mathrm{clust}}.
\end{equation}

To overlay the equipotential curves on the observational data, we evaluated $\Phi_{\mathrm{tot}}$ on a 2D spatial grid matching the field of view of the WISE image. The Galactic cylindrical coordinates $(R, Z)$ were computed by projecting the local offsets in Galactic longitude and latitude ($\Delta \ell \cos b$, $\Delta b$) on the sky plane, adopting a distance to Wd1 of \rev{$d = 4.21$\,kpc} and an oblique viewing angle $\theta = 20^\circ$. 

The white contours in Fig.~\ref{fig:wise_overlay} represent the equipotential levels, tracing the gravitational well of the system. These curves exhibit a clear North-South asymmetry, appearing more compressed towards the bottom of the map (negative $\Delta b$ in our projection), which reflects the vertical gradient of the Miyamoto-Nagai disk potential. 
Specifically, since Wd1 is located below the Galactic plane ($Z \approx \rev{-29.5$\,pc)}, moving toward the bottom of the map ($\Delta b < 0$) increases the absolute distance from the Galactic mid-plane (reaching $Z \approx -44.6$\,pc at the lower edge). Within the disk scale height, the vertical restoring force of the Milky Way increases with distance from the plane, creating a steeper background potential gradient that compresses the equipotential lines in the southern region. 

Conversely, toward the top of the map ($\Delta b > 0$), the system approaches the Galactic mid-plane ($Z \approx -14.6$\,pc at the upper edge), where the absolute vertical Galactic force is weaker, resulting in a lower potential slope and wider contour spacings. This structural asymmetry implies that the total gravitational confinement of the cluster is less restrictive in the northern direction. As a consequence, this shallower potential boundary facilitates the tidal detachment of the outer stellar layers, providing a natural explanation for the preferential accumulation of escaping stars on this side (Sect.~\ref{sec:unbound}).

\subsection{Galactic imprint on the Wd1 kinematics}

The spatial distribution of the identified cluster members overlaid on the mid-infrared emission (Fig.~\ref{fig:wise_overlay}) shows a striking morphological correspondence with the calculated effective gravitational equipotential contours. The system exhibits a pronounced North-South structural asymmetry, where the outermost stellar layers preferentially occupy the northern region of the cluster footprint ($\Delta b > 0$). This spatial configuration is likely due to the high reddening affecting the cluster. At the same time, it can also be indicative that the macroscopic structure of Wd1 is reflecting the environmental perturbations from the Galactic field.

This scenario has been tested with our 2D kinematic modeling. The clear dominance of the vertical velocity gradient ($\partial v_b/\partial b = 0.47 \pm 0.10~\mathrm{km\,s^{-1}\,pc^{-1}}$) over the radial and longitudinal components demonstrates that the internal kinematics of the cluster is associated to an out-of-plane, vertical apparent expansion mechanism (Sect.\,\ref{sec:kinematics}). Assuming an impulsive scenario where this collective expansion was triggered by a short-duration kinematic perturbation, the time elapsed since the event \rev{ ($\Delta t_{\mathrm{kin}}$)} can be  related to the velocity gradient by the relation \rev{$\Delta t_{\mathrm{kin}} = (\partial v_b/\partial b)^{-1}$.} By converting our empirical vertical gradient into temporal units, we derive an internal kinematic expansion timescale of $\Delta t_{\mathrm{kin}} = 2.08 \pm 0.44$\,Myr. 

To test whether this internal expansion timescale is physically connected to the macro-orbital history of the cluster within the MW, we performed a kinematic backtracking projection of Wd1's bulk motion. \rev{Given the current vertical position of the cluster center ($b = -0.4012^\circ$, corresponding to $Z \approx -29.5$\,pc at the  distance of $4.21$\,kpc) 
and our derived systemic vertical velocity ($v_b^0 = -13.74 \pm 0.30~\mathrm{km\,s^{-1}}$), 
we adopted the linear backtracking relation
$\Delta t_{\mathrm{orbit}}= |Z| / |v_z| \approx |Z| / |v_b^0|$ (where $v_z = v_b \cos b \approx v_b$ since $\cos b \approx 1$). This calculation reveals that the cluster crossed the Galactic mid-plane ($Z=0$, $b=0$) approximately $\Delta t_{\mathrm{orbit}}= |Z|/|v_b^0| \times 0.9778\,\mathrm{km\,s^{-1}}\mathrm{Myr\,\,pc^{-1}} = 2.10$\,Myr ago. }

\rev{To estimate the error due to observational astrometric uncertainties we consider 
that the kinematic crossing time is completely independent of the distance uncertainty  since both the vertical displacement $Z$ and the vertical velocity $v_b^0$ scale linearly with $d$.
The remaining observational error depends only on the precision of the astrometric proper motion fit.} 
\rev{Applying the standard error propagation 
we obtain:
$
\sigma_{\mathrm{obs}} = \Delta t_{\mathrm{orbit}}\times  \sigma_{v_b^0}/|v_b^0|
$.
Using our derived values of $\Delta t_{\mathrm{orbit}} = 2.10$~Myr and systemic vertical velocity $v_b^0 = -13.74 \pm 0.30~\mathrm{km\,s^{-1}}$, this yields a robust and tightly constrained observational error of $\sigma_{\mathrm{obs}} \approx 0.05$~Myr. }

\rev{Systematic contributions are negligible in comparison.
To quantify the impact of uncertainties in the adopted Galactic potential model, where non-linear
 orbit integration precludes a direct analytical propagation of errors, we performed a numerical 
 sensitivity analysis. 
 Assuming a hypothetical $\pm 10\%$ uncertainty in the background gravitational potential parameters \citep{mcmi17}, 
 we perturbed the parameters of the Miyamoto-Nagai disk ($M_d, a, b$) 
 and NFW halo ($\rho_s, R_s$) independently and evaluated the resulting 
 variations in the mid-plane crossing time through numerical integration of the vertical 
 equations of motion. The timing sensitivity is strictly dominated by the disk scale height
  ($\Delta T_b \approx 0.0054$\,Myr) and disk mass ($\Delta T_{M_d} \approx 0.0052$\,Myr),
   whereas halo parameters contribute negligibly ($\lesssim 0.0002$\,Myr). Therefore
 the Galactic potential parameters contributes $\sigma_{\rm potential} \approx 0.008$\,Myr, while the formal Gaia-based statistical uncertainty on the solar vertical displacement ($\sigma_{Z_\odot} \approx 0.3$\,pc; Bennett \& Bovy 2019) introduces a mid-plane reference uncertainty of $\sigma_{t, Z_\odot} \approx 0.02$\,Myr. Combining astrometric, potential, and mid-plane reference uncertainties in quadrature ($\sqrt{\sigma_{\rm obs}^2 + \sigma_{\rm potential}^2 + \sigma_{t, Z_\odot}^2}$) yields $\sigma_{\rm total} \approx \pm 0.05$\,Myr.}
%
%
%

%
%

The derived timescale is in excellent agreement with the independent orbital crossing timescale derived from VVVX data by \citet{orde26b}, further confirming the reliability of our linear backtracking approximation.


 The strong agreement between the independent internal kinematic expansion age 
($2.08 \pm 0.44$\,Myr) and the external orbital crossing time \rev{($2.10 \pm 0.05$\,Myr)}
provides a  temporal correlation, indicating that the onset of the 
out-of-plane expansion of Wd1 is contemporary with its recent Galactic disk crossing. 
Given the cluster's current age ($\sim 4-5$\,Myr), this $\sim 2$\,Myr timescale 
also matches the expected epoch for the first core-collapse supernovae from the 
most massive stellar progenitors.

\del{With the current data, it is not possible to unambiguously disentangle whether this 
anisotropic expansion was triggered by internal dynamical processes, such as 
binary disruptions or close stellar encounters, or if the local Galactic environmental 
field played a role during the disk crossing.} 

\rev{Interestingly, our derived stellar kinematic expansion timescale of $\Delta t_{\mathrm{kin}} \approx 2.08$\,Myr is  consistent with  the gaseous environment of the cluster. Based on H~I observations, \citet{koth07} identified a large expanding interstellar bubble surrounding Wd1 with a minimum dynamical age of $\sim 2.5$\,Myr. The temporal coincidence between the stellar kinematic expansion and the dynamical age of this H~I bubble strongly supports a scenario where a major energetic event, such as intense collective stellar winds from massive members or an early supernova explosion, evacuated the residual gas and simultaneously imparted an expanding kinematic signature on the outer stellar envelope.}

\rev{With the current observational data, it is not possible to unambiguously disentangle 
the underlying physical trigger of such expansion. It could be driven by internal processes, such 
as energetic feedback from early core-collapse supernovae, sudden gas expulsion, or core 
dynamical interactions, such as binary disruptions , close stellar encounters and hierarchical subcluster mergers \citep{kara24}, or 
by external Galactic environmental forces experienced during the mid-plane passage 
(such as gravitational tidal shocks). Therefore, rather than claiming a direct 
causal link, we present this temporal correlation as an empirical constraint 
for future investigations.}


The 103 unbound Wd1 members, which orbit under the Galactic potential
with a weak influence by the cluster potential, reflect this in their
spatial distribution
(Sect.\,\ref{sec:unbound}). These \del{escaping} \rev{unbound candidate} members are not distributed isotropically; instead, they are preferentially found in the northern outskirts of the cluster. This spatial asymmetry is perfectly consistent with the vertical structure of the Galactic potential. Toward the top of the map ($\Delta b > 0$ in Fig.\,\ref{fig:wise_overlay}), the system approaches the Galactic mid-plane, where the absolute vertical restoring force of the disk weakens, resulting in a shallower effective potential barrier and wider contour spacings. 

High-speed members moving northward encounter a less restrictive gravitational confinement, which directly facilitates their escape into the field. Kinematically, 
this asymmetric escape is intrinsically linked to the observed positive vertical 
velocity gradient ($\partial v_b/\partial b$). Because these escaping northern stars 
possess internal velocity vectors directed northward, they move slower than the 
cluster's downward systemic translation ($v_b^0 = \rev{-13.86} \pm 0.26~\mathrm{km\,s^{-1}}$). 
Consequently, in the absolute sky frame, these northern members exhibit a reduced 
net vertical speed compared to the more cohesive bulk motion of the core.
Conversely, within the central core where the cluster's self-gravity dominates, the bound stars exhibit internal velocities governed by the deep gravitational 
potential well. These members are  confined within the closed effective 
equipotential curves, as their short local dynamical timescales, characterized 
by a core crossing time of only $\sim 0.2$\,Myr \citep{genn17},
have allowed them 
to dynamically adjust and reach a state of quasi-equilibrium.

Finally, the physical properties of the unbound population provides  insight into the underlying escape mechanism. Traditional long-term stellar evaporation driven by two-body relaxation typically populates the escaping tail with stars possessing velocities very close to, or only marginally above, the local escape speed threshold ($v_{\mathrm{esc}}$). In 
contrast, the identified unbound members exhibit remarkably high excess speeds, reaching an average speed of 
$\langle v \rangle = \rev{18.06}~\mathrm{km\,s^{-1}}$
(with an observed dispersion of $\sigma_{v} = \rev{5.65}~\mathrm{km\,s^{-1}}$), 
which significantly exceeds the point-mass escape speed profile. After accounting for the astrometric uncertainties of \textit{Gaia} DR3 ($\sigma_{\mathrm{err}} \approx 3.27~\mathrm{km\,s^{-1}}$) by performing a subtraction in quadrature, we isolate a substantial, genuine intrinsic speed spread of $\sigma_{\mathrm{int}} \approx \rev{4.61}~\mathrm{km\,s^{-1}}$. This high-velocity tail is hardly compatible with a steady-state evaporation scenario. Furthermore, if these stars had been continuously evaporating since the disk crossing event $\sim 2$\,Myr ago at such high speeds, they would have already traveled a median distance of $\sim 37$\,pc, dispersing into an extended spatial envelope with an intrinsic structural width of $\pm 9.5$\,pc. Such a massive displacement would have placed the vast majority of these \del{escaping} \rev{unbound candidate} members completely outside our $13'$ field of view.

Their prominent presence within the immediate boundaries of the cluster footprint strongly indicates that these stars are the result of recent or ongoing dynamical ejection events. This scenario is directly confirmed by our kinematic point-back test.
\del{Specifically, we computed the alignment angle $\theta$ for each star, defined as the angle between its position vector relative to the cluster center and its velocity vector in the rest frame of the system.}
\rev{To evaluate whether these unbound stars are physical candidates for radially escaping runaways (or walkaways), we computed the projected alignment angle $\theta$ in the plane of the sky between the position vector $\vec{r}_{\mathrm{proj}} = (x, y)$ relative to the cluster center and the rest-frame physical velocity vector $\vec{v}_{\mathrm{2D}} = (v_l, v_b)$:
\begin{equation}
\cos \theta = \frac{x v_l + y v_b}{r_{\mathrm{proj}} \, v_{\mathrm{2D}}}
\end{equation}
where $\theta = 0^\circ$ corresponds to a purely radial outward trajectory on the sky plane. }

\rev{We emphasize that $\theta$ is explicitly a 2D projected angle on the sky plane. Because individual line-of-sight velocities ($v_{\mathrm{los}}$) are not available for the full faint sample, $v_{\mathrm{2D}}$ represents a lower limit to the true 3D spatial velocity ($v_{\mathrm{3D}} = \sqrt{v_{\mathrm{2D}}^2 + v_{\mathrm{los}}^2}$). Consequently, selecting stars with $\theta < 30^\circ$ specifically identifies candidates whose ejection trajectories lie predominantly near the sky plane. Stars ejected with large line-of-sight velocity components may project with larger 2D angles or lower projected speeds, making our identified sample of sky-projected runaway/walkaway candidates a conservative selection.}
Among the 103 unbound sources, we successfully isolate 14 candidate walkaway stars \citep{renz19}, that exhibit directional velocity vectors pointing radially outward from the core ($\theta < 30^\circ$, where $\theta = 0^\circ$ corresponds to a perfectly radial outflow) within a short kinematic timescale of $< 1.0$\,Myr. Such a population likely originated by energetic multi-body close encounters or the sudden disruption of tight binary systems in the dense cluster core, acting together with the ongoing influence by the Milky Way.
\subsection{Dynamical State of the Core }
\label{sec:core_dynamics}

Our derived intrinsic 1D velocity dispersion for the bound core ($\sigma_{\mathrm{int, core}} \approx 4.6 - 5.4~\mathrm{km\,s^{-1}}$, 
with a 1D quadratic mean of $\sim \rev{5.01}~\mathrm{km\,s^{-1}}$) provides crucial insights into the dynamical status of Wd1. This value is in  agreement with high-precision near-infrared and HST proper motion studies focused on the innermost, highly crowded core by \citet{wei25}, which report a central velocity dispersion of $4.13 \pm 0.13~\mathrm{km\,s^{-1}}$. The marginally higher dispersion derived here can be  attributed to the broader spatial coverage of our Gaia DR3 selection, which samples stars further out in the cluster potential where the radial velocity gradient and extended potential profile contribute to the overall spread.

To assess the virial state of the cluster core, we estimate the theoretical 1D virial velocity dispersion assuming a Plummer potential profile:
\begin{equation}
\sigma_{\mathrm{vir}} \approx \sqrt{\frac{G M_{\mathrm{tot}}}{6 r_c}} .
\end{equation}
Adopting a total cluster mass of 
$M_{\mathrm{tot}} = 4.4-5.7 \times 10^4\,\mathrm{M}_{\odot}$ \citep{ande17} and a core radius of $r_c \approx 1.0~\mathrm{pc}$, the expected virial dispersion is 
$\sigma_{\mathrm{vir}} = 5.6-7.0~\mathrm{km\,s^{-1}}$. 
This range is also consistent with the theoretical virial velocity dispersion of $\sigma_{\mathrm{vir}} = 4.5-6.5~\mathrm{km\,s^{-1}}$ estimated by \citet{wei25}.

Comparing the intrinsic dispersion value ($\sigma_{int,core} \approx 5.02~\mathrm{km\,s^{-1}}$) with the theoretical range ($\sigma_{\mathrm{vir}} \approx 5.6-7.0~\mathrm{km\,s^{-1}}$) shows that the observed intrinsic dispersion is slightly lower than the predicted virial limit ($\sigma_{\mathrm{int}} \le \sigma_{\mathrm{vir}}$). 
Therefore, we conclude that the cluster core resides in a sub-virial state/near virial equilibrium ($\alpha_{\mathrm{vir}} \le 1$), in  agreement with the conclusions of \citet{wei25}, who similarly inferred a bound sub-virial dynamical state for the central region of Wd1 based on a large sample of cluster members.

Conversely, the  unbound sample exhibits a significantly larger total velocity dispersion due to the inclusion of the escaping, high-velocity population ($\sigma_{\mathrm{int}} \approx 11 - 15~\mathrm{km\,s^{-1}}$). This dichotomy strongly reinforces our scenario of two kinematically decoupled subsystems: a gravitationally bound core in near equilibrium state, embedded within an expanding, unbound stellar envelope.

\section{Summary and conclusions}
\label{sec:conclusions}
In this work, we have performed a comprehensive 5D astrometric and kinematic study of the supermassive young cluster Wd1 using \textit{Gaia}~DR3 data. By applying the DBSCAN clustering algorithm to a multidimensional space (RA, Dec, $\mu_{\alpha*}$, $\mu_\delta$, and photo-geometric distance), we identified 559 high-probability member candidates of Wd1, located at a median distance of \rev{$4.21 \pm 0.20$}\,kpc. Specifically, this census represents a significant increase of 158 members compared to the 401 bona fide members previously identified by \citet{negu22}.
In addition, we identified
658 members of the foreground cluster BH197, located at a distance of $2.07 \pm 0.21$ \,kpc.

By using a clean sample of 524 Wd1 members with proper motion errors smaller than 0.5~mas~yr$^{-1}$, we performed a detailed cluster kinematic analysis. 
By computing the internal 2D speed $v$ for each star and comparing it with a theoretical escape speed profile $v_{\mathrm{esc}}(r) = \sqrt{2GM/r}$ (assuming a cluster mass of $5 \times 10^4\,\mathrm{M}_\odot$), we separated the sample into two distinct kinematic populations. By applying a strict selection criterion ($v - \sigma_{v,i} > v_{\mathrm{esc}}$) to account for individual astrometric uncertainties, we successfully disentangled a  core population of 421 bound members from a robust sub-population of 103 unbound stars actively escaping the system.
Of these, 14 are classified as walkaway stars.

In addition, we modeled the internal velocity field as superposed translation, expansion and rotation components.
 We derived a precise systemic bulk motion for the cluster \del{core} of $\mu_{l*} = -4.26 \pm 0.02$~mas~yr$^{-1}$ and $\mu_b = -0.70 \pm 0.02$~mas~yr$^{-1}$ in Galactic coordinates
($\mu_{\alpha*} = -2.25 \pm 0.02$~mas~yr$^{-1}$ and $\mu_\delta = -3.68 \pm 0.02$~mas~yr$^{-1}$ in equatorial coordinates),
corresponding to a horizontal velocity of $v_l^0 = \rev{-85.07}\pm 0.38~\mathrm{km\,s^{-1}}$ and a vertical velocity of $v_b^0 = \rev{-13.74} \pm 0.30~\mathrm{km\,s^{-1}}$.
 These values are in excellent agreement with near-infrared measurements from the VVVX survey
 ($\mu_{\alpha*} =-2.36 \pm 0.02$~mas~yr$^{-1}$ and $\mu_\delta = -3.70 \pm 0.02$~mas~yr$^{-1}$ in equatorial coordinates) derived by \citet{orde26b}, providing a strong cross-validation of our membership census. 
 
 We detected a statistically significant anisotropic apparent expansion ($>4\sigma$) primarily directed orthogonal to the Galactic plane, with the principal expansion axis corresponding to $PA_{\mathrm{gal}} = 164.0^\circ\pm\rev{10.1}^\circ$. In addition, the analysis revealed a strong cross-gradient ($\partial v_l / \partial b$), which provides a clear kinematic signature of the tidal torque exerted by the Galactic potential as the cluster resides very close to the Galactic thin disk. 

Our kinematic results provide an interesting physical framework when correlating the cluster's internal dynamics with its large-scale Galactic orbit.
By converting our empirical vertical velocity gradient into temporal units, we derived a data-driven kinematic expansion timescale of $\Delta t_{\mathrm{kin}} = 2.08 \pm 0.44$\,Myr. A simple linear backtracking of the cluster's systemic motion independently demonstrates that Wd1 crossed the Galactic mid-plane ($b=0$) \rev{$\Delta t_{\mathrm{orbit}} \approx 2.10 \pm 0.05$\,Myr ago}. \del{The agreement between these two independent internal and external timescales 
presents an interesting temporal coincidence, placing the onset of the vertical 
expansion field at an epoch contemporary with the cluster's recent passage 
through the Galactic disk.}
\rev{The close agreement between these independent internal and external timescales reveals a tight temporal alignment. However, current data cannot unambiguously identify the underlying trigger, which could stem either from internal dynamical processes (e.g., early supernova feedback or binary disruptions) or from external Galactic environmental influences during the disk passage.
}

This scenario is also validated by our modeling of the local effective Galactic potential ($\Phi_{\mathrm{tot}}$). The background equipotential curves display a clear vertical asymmetry, becoming tighter at negative $\Delta b$ and widely spread toward the top of the map ($\Delta b > 0$) closer to the Galactic plane. This shallower gravitational boundary perfectly matches the observed spatial distribution of the unbound population, whose escape has been favored by the 
lower potential in the north regions where they are preferentially found.

Kinematic analysis reveals that the cluster core is in a gravitationally bound, sub-virial state ($\sigma_{\mathrm{int}} \approx \rev{5.01}~\mathrm{km\,s^{-1}} \le \sigma_{\mathrm{vir}}$), kinematically decoupled from an expanding, highly dispersed unbound envelope ($\sigma_{\mathrm{int}} \approx 11-15~\mathrm{km\,s^{-1}}$).
\begin{acknowledgements}
\rev{We thank the anonymous referee for useful comments and suggestions that helped improve the quality of the manuscript.}
This research was supported by the International Space Science Institute (ISSI) in Bern,
through ISSI International Team project \#25-639. 
L.P. acknowledges support from the Italian Ministero dell'Università e della Ricerca and
the European Union - Next Generation EU through project PRIN 2022 PM4JLH ``Know your little neighbours: characterising low-mass stars and planets in the Solar neighbourhood''.
FD and LP acknowledge partial support from INAF Mini-Grant 2023 Program PI: Damiani.
R.B. acknowledges support from Minigrant INAF PACE PI: Bonito (1.05.23.04.02)
M.T. acknowledges support from the Targeted Research Funds 2026 (FFR 2026) of the University of Palermo. 
M.T. acknowledges support from the PRIN 2022: Spatio-temporal Functional Marked Point Processes for probabilistic forecasting of earthquakes 2022BN7CJP P. I. Giada Adelfio. CUP B53C24006340006.
V.A-A and M.G.G. acknowledge support from the INAF grants 1.05.12.05.03 and 1.05.24.07.02.
K.M. acknowledges support from the  Fundação para a Ciência e a Tecnologia (FCT) through the CEEC-individual contract 2022.03809.CEECIND, and grant UID/04434/2023.
IN is partially supported by the Spanish Government Ministerio de Ciencia, Innovaci\'on y Universidades and Agencia Estatal de Investigación (MCIU/AEI/10.130 39/501 100 011 033/FEDER, UE) under grant PID2024-159329NB-C22.
This work presents results from the European Space Agency (ESA) space mission Gaia. Gaia data are being processed by the Gaia Data Processing and Analysis Consortium (DPAC). Funding for the DPAC is provided by national institutions, in particular the institutions participating in the Gaia MultiLateral Agreement (MLA). The Gaia mission website is \url{https://www.cosmos.esa.int/gaia}. The Gaia archive website is \url{https://archives.esac.esa.int/gaia}.
\end{acknowledgements}
\bibliography{wd1bib}{}
\bibliographystyle{aa}
\begin{appendix} 
\nolinenumbers
\section{Identification of the foreground cluster BH197}
\label{app:bh197}
Following the characterization of the main cluster, we applied the same 5D DBSCAN clustering methodology to the photometric sample associated with the foreground population. For this subset, we introduced an additional quality constraint on the parallax significance ($\varpi/\sigma_{\varpi} \geq 5$) to ensure the reliability of the astrometric signal for these closer sources, resulting in an input sample of 3027 stars.

The clustering parameters were optimized to $\epsilon=0.6$ and \texttt{min\_samples}=70. This analysis successfully isolated a second coherent group, corresponding to the foreground open cluster BH197 already identified in \citet{negu22}.

\begin{figure*}
   \centering
   \includegraphics[width=0.95\textwidth]{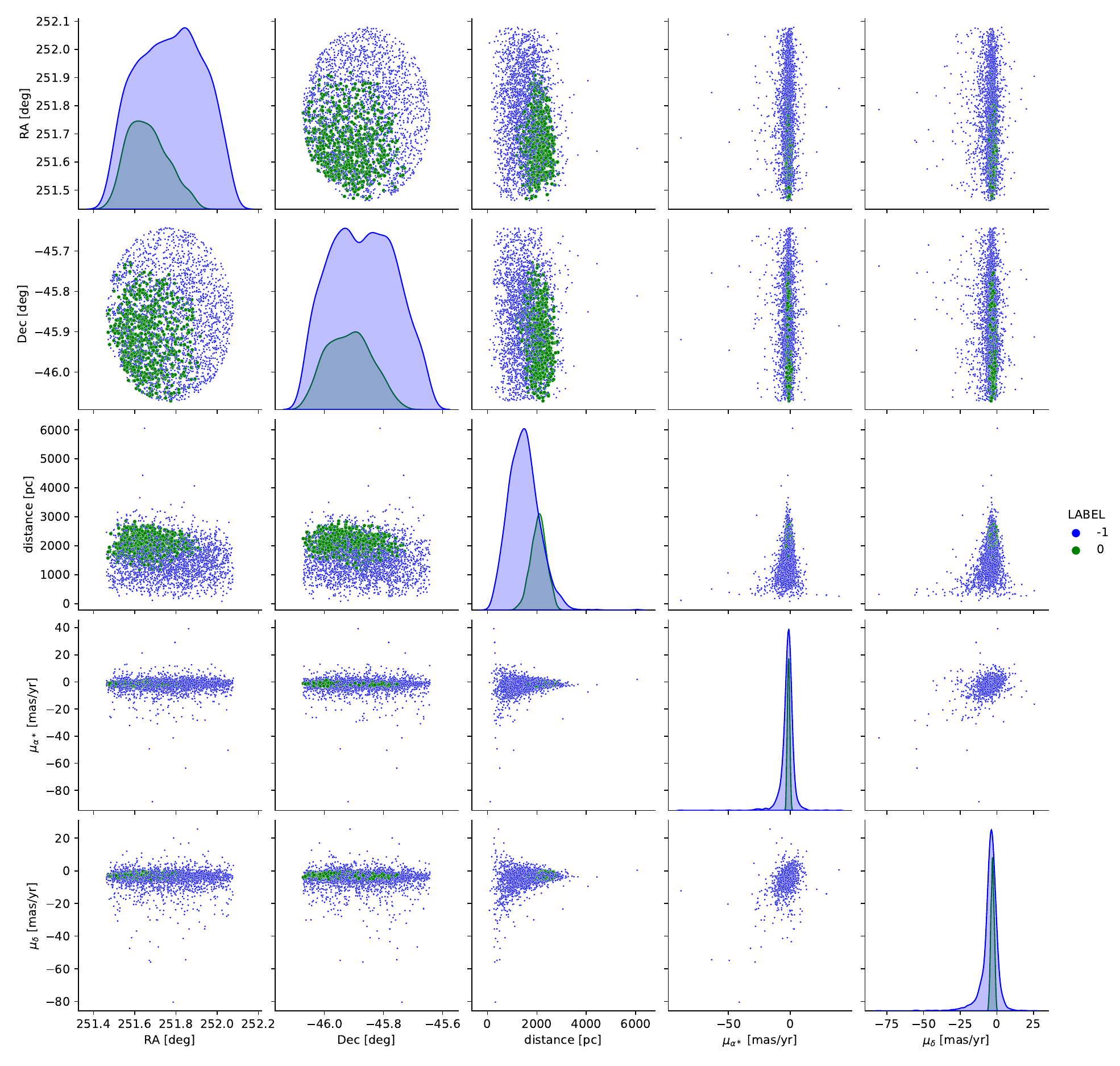}
   \caption{DBSCAN clustering results for the foreground open cluster BH197. The pair plot displays the distribution of the 658 candidate members (green points) identified within the 5D astrometric parameter space out of an initial sample of 3027 stars. Field stars, classified as noise by the algorithm, are represented by blue points. The Kernel Density Estimation (KDE) plots are shown along the diagonal.}
   \label{fig:bh197_dbscan}
\end{figure*}

The algorithm identified 658 high-probability members for BH197, which exhibit distinct kinematic and distance properties compared to Wd1. Specifically, they are concentrated at a median distance of $2.07$ kpc (with a median absolute deviation of $\sim 0.21$ kpc) and possess proper motions that are clearly offset from those of the Wd1 members. As illustrated in the pair plot in Fig. \ref{fig:bh197_dbscan}, while BH197 and Westerlund 1 partially overlap in sky projection ($\alpha, \delta$), they are fully resolved as separate entities in the 5D astrometric space, driven primarily by their distinct distances and proper motion vectors.

The complete catalog of the 658 selected members of BH197, including their \textit{Gaia} DR3 astrometric and photometric parameters alongside the photo-geometric distances, is provided in Table~\ref{tab:bh197_catalog}.
\input{FIGURES/bh197_mem}
\end{appendix} 
\end{document}

%% file: FIGURES/wd1_mem.tex
\begin{table*}[htbp]
\caption{Astrometric, photometric, distance, membership probability, and kinematic parameters for the selected \textit{Gaia} DR3 members of Westerlund 1.}
\label{tab:w1_catalog}
\centering
\resizebox{\textwidth}{!}{
\begin{tabular}{cccccccccccc}
\hline\hline
Source ID & $\alpha$ & $\delta$ & $\varpi$ & $\mu_{\alpha*}$ & $\mu_\delta$ & $G$ & $G_{BP}$ & $G_{RP}$ & $d_{\rm photogeo}$ & $p_{\rm MC}$ & $\text{Flag}_{\rm kin}$ \\
 & (deg) & (deg) & (mas) & (mas/yr) & (mas/yr) & (mag) & (mag) & (mag) & (pc) & & \\
\hline
5943100939338998272 & 251.53807 & -45.97700 & $0.154 \pm 0.040$ & $-2.71 \pm 0.05$ & $-3.57 \pm 0.05$ & 13.222 & 15.821 & 11.798 & $3888.1_{-428.8}^{+598.3}$ & $0.975 \pm 0.005$ & 1 \\
5943100973698746880 & 251.53444 & -45.96164 & $0.161 \pm 0.118$ & $-2.34 \pm 0.15$ & $-2.77 \pm 0.13$ & 14.733 & 19.683 & 13.083 & $3653.6_{-1016.5}^{+951.4}$ & $0.848 \pm 0.011$ & 1 \\
5943101042399178752 & 251.57560 & -45.95595 & $0.086 \pm 0.276$ & $-3.35 \pm 0.36$ & $-3.59 \pm 0.31$ & 18.383 & 21.367 & 16.517 & $3985.5_{-852.8}^{+1353.9}$ & $0.644 \pm 0.015$ & 1 \\
5943101072457361024 & 251.55073 & -45.95928 & $0.064 \pm 0.126$ & $-2.67 \pm 0.16$ & $-3.10 \pm 0.14$ & 17.895 & 20.507 & 16.479 & $4606.3_{-1722.1}^{+2389.4}$ & $0.688 \pm 0.015$ & 1 \\
5943101076777961344 & 251.54740 & -45.95512 & $0.172 \pm 0.076$ & $-2.50 \pm 0.09$ & $-4.00 \pm 0.09$ & 16.425 & 19.330 & 14.956 & $4539.3_{-1182.6}^{+1941.1}$ & $0.813 \pm 0.012$ & 1 \\
\hline
\end{tabular}
}
\tablefoot{Only the first 5 rows are shown as guidance. The full catalog is available in electronic form at the CDS. Uncertainties correspond to $1\sigma$ measurement errors. Col. (11): Empirical Monte Carlo membership probability ($p_{\text{MC}}$) and its binomial standard error derived from $B=1000$ iterations (Sect. 3.3). Col. (12): Kinematic state flag ($\text{Flag}_{\text{kin}}$), where \texttt{0} denotes a gravitationally bound star, \texttt{1} indicates a candidate unbound cluster member, and \texttt{2} highlights an unbound star classified as a walkaway/runaway candidate (Sect. 4.1).}
\end{table*}

%% file: FIGURES/taylor_exp_fit_all_samples_spatial_errpa_noerr.tex
\begin{table*}
\caption{\rev{Results of the 2D kinematic linear fit for different subsamples of Westerlund 1. Physical units are computed assuming $d = 4.21$ kpc.}}
\label{tab:kinematics_all_samples}
\centering
\resizebox{\textwidth}{!}{
\begin{tabular}{l c c c c c c c c c c c}
\hline\hline
Sample & N & $v_l^0$ & $\partial v_l / \partial l$ & $\partial v_l / \partial b$ & $v_b^0$ & $\partial v_b / \partial l$ & $\partial v_b / \partial b$ & Div & Vort & PA$_{\rm gal}$ & $\chi^2_\nu(l, b)$ \\
 & & (km s$^{-1}$) & (km s$^{-1}$ pc$^{-1}$) & (km s$^{-1}$ pc$^{-1}$) & (km s$^{-1}$) & (km s$^{-1}$ pc$^{-1}$) & (km s$^{-1}$ pc$^{-1}$) & (km s$^{-1}$ pc$^{-1}$) & (km s$^{-1}$ pc$^{-1}$) & (deg) & \\
\hline
Bound & 421 & $-85.14 \pm 0.28$ & $-0.13 \pm 0.11$ & $-0.15 \pm 0.13$ & $-13.86 \pm 0.26$ & $-0.21 \pm 0.10$ & $0.27 \pm 0.11$ & $0.14 \pm 0.16$ & $-0.06 \pm 0.16$ & $159.2 \pm 8.5^\circ$ & 11.3, 5.4 \\
DBSCAN All (Bound+Unbound) & 524 & $-85.07 \pm 0.38$ & $0.11 \pm 0.10$ & $0.19 \pm 0.12$ & $-13.74 \pm 0.30$ & $-0.41 \pm 0.08$ & $0.47 \pm 0.10$ & $0.58 \pm 0.14$ & $-0.60 \pm 0.15$ & $164.0 \pm 10.1^\circ$ & 26.2, 15.7 \\
Unbound & 103 & $-85.56 \pm 1.86$ & $0.32 \pm 0.24$ & $0.56 \pm 0.35$ & $-14.01 \pm 1.29$ & $-0.53 \pm 0.16$ & $0.62 \pm 0.24$ & $0.94 \pm 0.34$ & $-1.09 \pm 0.39$ & $2.3 \pm 33.1^\circ$ & 87.8, 60.0 \\
Field & 322 & $-81.83 \pm 3.64$ & $-0.00 \pm 0.40$ & $0.41 \pm 0.56$ & $-8.83 \pm 2.17$ & $-0.02 \pm 0.24$ & $-0.36 \pm 0.33$ & $-0.37 \pm 0.52$ & $-0.43 \pm 0.61$ & $66.3 \pm 30.8^\circ$ & 7391.9, 1192.4 \\
Bound + Field (MC 1000x) & 521 & $-84.47 \pm 0.99$ & $-0.07 \pm 0.25$ & $0.36 \pm 0.29$ & $-13.31 \pm 0.62$ & $-0.15 \pm 0.16$ & $0.15 \pm 0.18$ & $0.07 \pm 0.31$ & $-0.49 \pm 0.33$ & $48.7 \pm 18.8^\circ$ & 856.5, 170.0 \\
\hline
\end{tabular}
}
\tablefoot{Divergence (Div) and Vorticity (Vort) map the cluster's isotropic expansion and rigid rotation, respectively. PA$_{\rm gal}$ indicates the Position Angle (measured East of Galactic North) of the principal axis of maximum expansion. The $\chi^2_\nu$ column provides the reduced Chi-squared for the $(l, b)$ axes, acting as a proxy for the internal kinematic coherence of each sample.}
\end{table*}

%% file: FIGURES/bh197_mem.tex
\begin{table*}[htbp]
\caption{Astrometric, photometric, and distance parameters for the 658 selected \textit{Gaia} DR3 candidate members of BH\,197.}
\label{tab:bh197_catalog}
\centering
\resizebox{\textwidth}{!}{
\begin{tabular}{cccccccccc}
\hline\hline
Source ID & $\alpha$ & $\delta$ & $\varpi$ & $\mu_{\alpha*}$ & $\mu_\delta$ & $G$ & $G_{BP}$ & $G_{RP}$ & $d_{\rm photogeo}$ \\
 & (deg) & (deg) & (mas) & (mas/yr) & (mas/yr) & (mag) & (mag) & (mag) & (pc) \\
\hline
5943100286503950080 & 251.56914 & -46.02052 & $0.611 \pm 0.048$ & $-2.14 \pm 0.06$ & $-3.28 \pm 0.05$ & 16.302 & 17.171 & 15.379 & $1577.4_{-113.8}^{+138.4}$ \\
5943100591420977152 & 251.59513 & -46.02174 & $0.506 \pm 0.085$ & $-1.99 \pm 0.11$ & $-4.34 \pm 0.09$ & 17.362 & 18.154 & 16.515 & $1993.7_{-275.8}^{+413.7}$ \\
5943100595741593600 & 251.59948 & -46.02145 & $0.507 \pm 0.022$ & $0.13 \pm 0.03$ & $-0.98 \pm 0.02$ & 14.494 & 15.080 & 13.749 & $1821.8_{-99.6}^{+78.3}$ \\
5943100625780728576 & 251.60472 & -46.01161 & $0.473 \pm 0.065$ & $0.17 \pm 0.09$ & $-1.40 \pm 0.07$ & 16.803 & 17.470 & 16.010 & $2103.5_{-228.6}^{+346.9}$ \\
5943100625780731776 & 251.59021 & -46.00788 & $0.824 \pm 0.136$ & $-0.75 \pm 0.18$ & $-1.77 \pm 0.15$ & 18.169 & 19.021 & 17.340 & $1590.4_{-214.7}^{+527.7}$ \\
\hline
\end{tabular}
}
\tablefoot{Only the first 5 rows are shown as guidance. The full catalog is available in electronic form at the CDS. Uncertainties correspond to $1\sigma$ measurement errors. Distances are the photo-geometric estimates from \citet{bail21}.}
\end{table*}